\documentclass[sigconf,nonacm]{acmart}
\setcopyright{cc}
\setcctype{by}
\usepackage{makecell}
\begin{document}

\title[Four Types of LLM Reliance and Their Predictors Among Undergraduate Writers]{Four Types of LLM Reliance and Their Predictors Among Undergraduate Writers}
\subtitle{A Mixed-Methods Study at a Minority-Serving R1 University}

\author{Shahin Hossain}
\orcid{0000-0002-3461-1147}
\affiliation{%
  \institution{School of Education, University of Maryland, Baltimore County}
  \city{Baltimore}
  \state{MD}
  \postcode{21250}
  \country{USA}}
\email{shahinh1@umbc.edu}

\renewcommand{\shortauthors}{Hossain}

\begin{abstract}
Although most undergraduates now use Large Language Models (LLMs) for academic writing, there are no clear or validated methods to distinguish among the various ways students rely on AI. Existing tools assess AI reliance exclusively by frequency of use, a measure that, as this study demonstrates, inadvertently rewards students for depending on AI rather than recognizing their individual intellectual contributions. Conducted at a public minority-serving university and grounded in the AI Literacy Framework, Expectancy-Value Theory, and Biggs's 3P Model, the research involved 382 undergraduates, 14 interviews, and 396 open-ended responses from two survey items. Four distinct types of AI reliance were identified and confirmed: Strategic (34.3\%), Instrumental (30.9\%), Dialogic (30.4\%), and Dependent (4.5\%). Students' beliefs about the value and difficulty of using AI affect the extent of their reliance on AI, while their actual AI skills determine the specific type of AI they use, indicating that differentiated approaches are necessary to support students effectively. A key finding was that Strategic users, those who use AI most thoughtfully, scored the lowest on standard assessment measures, a discrepancy that reflects limitations in current evaluation methods, which prioritize AI-driven outcomes over actual writing quality and thereby penalize students demonstrating greater independent thinking. Analysis also revealed an additional group, approximately 13 percent, who decline AI use for ethical reasons rather than practical considerations, and whom current frameworks fail to account for. These findings carry significant implications for AI literacy programs, the measurement of student learning outcomes, and equitable AI policies at minority-serving institutions.
\end{abstract}

\keywords{LLM reliance, AI literacy, expectancy-value theory, undergraduate writing, minority-serving institutions, mixed-methods}

\maketitle

\section{Introduction}

The public release of ChatGPT in November 2022 significantly altered undergraduate academic writing practices, surpassing the capacity of institutional frameworks to respond effectively. ChatGPT, a generative artificial intelligence (GenAI) model developed by OpenAI, is ``specifically designed to generate human-like text in a conversational style'' \citep[p.~228]{cotton2024}. Within three academic years of its introduction, survey data show that approximately 88\% of undergraduates reported using AI tools in their assessments \citep{freeman2025}. Nevertheless, pedagogical, assessment, and policy frameworks governing such use have remained largely static. Academic research has paralleled the rapid adoption of these technologies, producing an expanding body of literature on prevalence, academic integrity risks, student perceptions, and the ethics of AI-generated text \citep{cotton2024,ng2021}. Despite this proliferation, the literature lacks a theoretically grounded analysis of what it means, in qualitatively distinct terms, to rely on large language models (LLMs) during academic writing. The discipline learned to count how much students use these tools long before it learned to distinguish the kinds of use that deepen thought from those that quietly replace it.

The limitation is structural. Existing research conceptualizes reliance on LLMs as a single, continuous frequency construct, categorizing students as users or non-users, or as frequent or infrequent users, and measuring outcomes by the extent to which GenAI contributes to their work. This approach obscures distinctions that are educationally significant. For instance, two students may submit the same research paper. One student may use GenAI to verify citations, assess the coherence of a drafted paragraph, and challenge claims about which she is uncertain. Another student may begin the assignment by prompting the AI and constructing the paper's intellectual framework based on the AI's output. According to all currently published measurement frameworks, these students are distinguished, if at all, only by frequency counts on a Likert scale. However, their cognitive engagement, developmental trajectories as writers, metacognitive regulation, and academic integrity practices differ fundamentally. Reliance on frequency measures alone cannot capture these distinctions. Frequency is the easiest property of reliance to measure and the least important to understand.

The empirical consequences of this gap are increasingly apparent. Recent experimental studies demonstrate substantial cognitive costs associated with AI-assisted writing. For instance, students who utilize LLMs exhibit reduced neural connectivity and experience difficulty recalling content from essays they have just composed, a pattern referred to as accumulating ``cognitive debt'' \citep[p.~151]{kosmyna2025}. Furthermore, AI-assisted writers engage in significantly fewer metacognitive activities compared to their unaided counterparts, a phenomenon termed ``metacognitive laziness'' \citep[p.~492]{fan2025}. Frequent AI use is also negatively correlated with critical thinking, as increased cognitive offloading reduces students' ``engagement in deep, reflective thinking'' \citep[p.~2]{gerlich2025}. Although these findings are noteworthy, they share a common methodological limitation: each study compares users with non-users without a framework for distinguishing qualitatively different forms of AI engagement. Cognitively preserving AI engagement, which maintains metacognitive oversight and authorial control, is analytically indistinguishable from engagement that results in cognitive debt unless the measurement instrument is specifically designed to capture this distinction. A field that cannot distinguish cognitively preserving engagement from cognitive abdication will, in time, mistake the one for the other and build its evidence base upon the confusion.

This research addresses the distinction by utilizing the AI Literacy Framework \citep[p.~1]{long2020,ng2021}, Expectancy-Value Theory \citep{wigfield2000}, and Biggs' \citeyearpar{biggs1993b} Presage-Process-Product Model. Using a sequential explanatory mixed-methods design with 382 undergraduates at a public minority-serving research university, the study identifies and validates four empirically distinct types of LLM reliance: Strategic, Instrumental, Dialogic, and Dependent. It further models the individual-difference predictors that determine both the type and intensity of reliance adopted by students. The research makes three primary contributions to the literature. First, it introduces a theoretically grounded and empirically validated typology that distinguishes qualitatively different forms of LLM reliance, thereby addressing the conceptual limitations of frequency-based frameworks. Second, it demonstrates that two distinct predictive systems govern reliance: AI literacy determines type membership, while Expectancy-Value beliefs determine intensity. These systems require separate and non-interchangeable interventions. Third, the study identifies a fundamental validity issue in existing LLM outcome measurement, the resolution of which is essential for advancing interpretive progress in the field. Why students rely on AI and how they rely on it answer to different forces, and an intervention aimed at one will not move the other.

The third contribution warrants explicit articulation at the outset. The most theoretically significant finding of this study is not the typology itself, but rather the observation that Strategic users, those who are deliberate, critically evaluative, and academically principled, consistently score lowest on all outcome measures. A superficial interpretation may suggest that principled AI restraint is less effective than uncritical reliance. However, structural analysis indicates that existing outcome instruments in the LLM literature predominantly assess statements such as ``I use generative AI tools to achieve X,'' thereby measuring AI usage rather than writing quality or cognitive development. Students who intentionally restrict AI involvement in their work score low on these measures by design, not due to weak writing, but because their principled restraint results in lower AI-attributed achievement. Qualitative data in this study support this reinterpretation. What has been labeled as ``outcomes'' in the field is, at best, an indicator of AI's contribution to student work and, at worst, a metric that systematically rewards dependence while mis-characterizing the most cognitively sophisticated students as the weakest performers. An instrument that scores the magnitude of AI's contribution will, by its own logic, rank the most self-reliant writer beneath the most dependent one and report the inversion as an outcome. In such a measure, principled restraint becomes indistinguishable from incapacity.

The institutional context of this study amplifies its significance. The research was conducted at a public R1 minority-serving institution (MSI) and Asian American and Native American Pacific Islander-Serving Institution (AANAPISI) in the mid-Atlantic, where over half of undergraduates identify as members of minority groups and approximately one-third are first-generation college students. This study investigates LLM reliance in a setting where its consequences are especially pronounced and unevenly distributed. At MSIs, structural heterogeneity in academic preparation leads to disparities in AI literacy, which in turn exacerbate existing inequities. Students with greater prior AI literacy are better positioned to use LLMs to support cognitive development, while those with limited literacy may accept AI-generated output without critical evaluation, resulting in superficially polished work that does not enhance underlying capabilities. First-generation status emerged as a significant predictor of reliance intensity, indicating a compensatory dynamic in which AI supplements academic resources that are otherwise more accessible to advantaged peers. These equity considerations are central to the study's findings.

This paper is structured as follows. The Literature Review synthesizes research on LLMs in academic writing, writing development theory, equity dimensions of the GenAI literacy divide, and academic integrity in the post-plagiarism era. The Theoretical Framework section integrates three explanatory frameworks and delineates a four-type taxonomy. The Method section outlines the sequential explanatory mixed-methods design. The Results section addresses each research question in sequence. The Discussion situates the findings within the broader literature and elaborates on their practical and theoretical implications.

\section{Theoretical Framework}

\begin{figure}
\centering
\includegraphics[width=\linewidth]{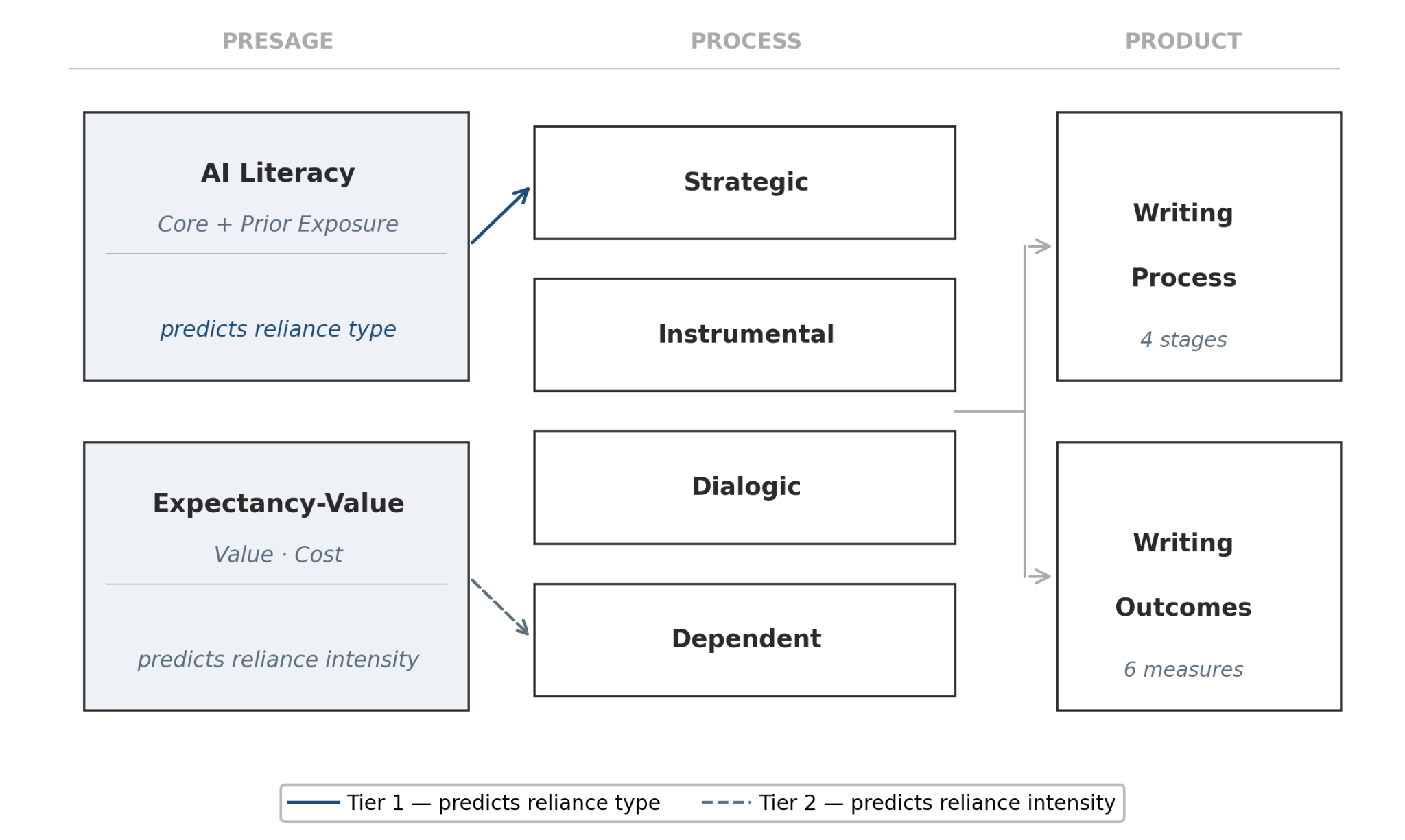}
\caption{Two-Tier Predictive Architecture: AI Literacy (Type Direction) $\times$ EVT (Intensity) $\times$ 3P Model (Process Structure)}
\Description{Diagram showing the two-tier predictive architecture linking AI literacy and expectancy-value beliefs (presage) to four reliance types (process) and writing outcomes (product).}
\label{fig:architecture}
\end{figure}

This study integrates three complementary frameworks that operate at two distinct explanatory levels. At the predictive level, the AI Literacy Framework \citep{long2020} accounts for the \emph{type} of reliance students adopt, representing the qualitative direction of engagement. In contrast, Expectancy-Value Theory \citep[EVT;][]{wigfield2000} explains the \emph{intensity} with which students rely on large language models (LLMs), reflecting the quantitative magnitude of engagement. At the structural level, Biggs' \citeyearpar{biggs1993b} Presage-Process-Product (3P) Model provides the architectural scaffolding that connects student characteristics (presage), reliance behavior (process), and writing outcomes (product) within a unified causal system. Collectively, these frameworks form a two-tier predictive architecture, comprising separate yet interlocking explanatory systems, as illustrated in Figure~\ref{fig:architecture}. The central claim supported by this architecture is that AI literacy and EVT beliefs function as distinct, non-redundant predictors: a student may possess high EVT beliefs and adopt any of the four reliance types, while a student with high AI literacy may still rely intensively but will do so strategically. Distinct mechanisms govern different dimensions of engagement, and interventions targeting only one dimension will not affect the other.

\subsection{The AI Literacy Framework: Predicting Reliance Type}

\citeauthor{long2020}'s \citeyearpar{long2020} AI Literacy Framework defines AI literacy as a set of competencies that enable individuals to understand how AI functions, critically evaluate AI technologies, and communicate and collaborate effectively with AI (as corroborated by \citealp[p.~4]{ng2021}; \citealp[p.~4]{allen2024}). Ethical reasoning, which is central to the present study, is emphasized more explicitly in subsequent educational frameworks (\citealp[p.~1]{ng2021}; \citealp[p.~7]{allen2024}). \citet{ng2021} identify four dimensions of AI literacy: ``know and understand, use and apply, evaluate and create, and ethical issues'' (p.~1). In this study, these are reframed as knowing and understanding AI, using and applying AI, evaluating and creating AI, and navigating AI ethics. The first three dimensions correspond to ascending cognitive levels in Bloom's taxonomy (p.~4; Fig.~2, p.~5), while AI ethics is treated as a cross-cutting consideration. \citeauthor{allen2024}'s \citeyearpar{allen2024} ED-AI Lit framework includes six components: Knowledge, Evaluation, Collaboration, Contextualization, Autonomy, and Ethics (p.~3). For the purposes of this study, three conditions are emphasized: Knowledge, Evaluation, and Ethics (pp.~4--7). These conditions are posited as prerequisites for strategic, rather than uncritical, engagement with LLMs: sufficient conceptual knowledge of LLM functionality and limitations; developed skills for evaluating and verifying AI-generated outputs; and ethical awareness regarding the consequences of academic AI use.

Importantly, higher AI literacy does not reduce engagement with AI but instead redirects it. \citet{zhai2024}, in a systematic review, found that over-reliance on AI dialogue systems is associated with diminished decision-making, critical thinking, and analytical reasoning. However, this risk is not uniform; higher AI literacy shifts engagement toward verification and critical evaluation rather than indiscriminate offloading, enabling students with greater literacy to utilize AI tools in ways that preserve and extend cognitive engagement. This redirecting function constitutes the mechanism by which AI literacy predicts reliance type: students with higher literacy are more likely to adopt Strategic or Dialogic reliance and less likely to adopt Dependent reliance. In this study, AI literacy is measured using two subscales: Core AI Literacy (six items assessing knowledge, evaluation, and ethical awareness; $\alpha = .775$) and Prior Exposure (seven items capturing breadth of experience; $\alpha = .664$), following \citeauthor{allen2024}'s \citeyearpar{allen2024} educational extension.

\subsection{Expectancy-Value Theory: Predicting Reliance Intensity}

Expectancy-Value Theory \citep[pp.~68--69]{wigfield2000} posits that behavioral engagement is governed by two psychological constructs: \emph{expectancy for success}, defined as the student's belief in their ability to perform effectively, and \emph{subjective task value}, which reflects the perceived usefulness, enjoyment, or importance of the task. When applied to LLM engagement, EVT predicts that students who believe they can use AI effectively and who attribute high utility, attainment, or intrinsic value to AI assistance will rely on LLMs more extensively and persistently than those with lower expectancy or value beliefs. The theory's \emph{cost} dimension, described by \citet{wigfield2000} as the extent to which engaging in one activity ``limits access to other activities,'' along with the required effort and its emotional cost (p.~72), is particularly relevant: students who perceive the cognitive cost of AI engagement as low are likely to rely more broadly, regardless of their literacy level. \citeauthor{fan2025}'s \citeyearpar{fan2025} randomized experiment provides convergent evidence that ChatGPT support ``may promote learners' dependence on technology and potentially trigger metacognitive `laziness'\,'' (p.~490), thereby improving short-term essay performance without corresponding gains in knowledge transfer. Although Fan et al.\ interpret this through the lens of cognitive offloading, where learners ``delegate cognitive tasks to external tools to reduce cognitive effort'' (p.~492), the dynamic aligns with EVT's cost mechanism: when AI reduces the perceived effort cost of a task, learners are more likely to rely on it rather than engage deeply, even at the expense of metacognitive regulation that supports transfer. This distinction clarifies EVT's predictive role relative to AI literacy: EVT explains the extent of students' reliance on it, while AI captures the manner in which they rely on it. EVT is operationalized with five items assessing expectancy for success, efficiency value, intrinsic enjoyment, attainment value, and utility value ($\alpha = .923$).

\subsection{Biggs' Presage-Process-Product Model}

Biggs' \citeyearpar{biggs1993b} Presage-Process-Product (3P) Model conceptualizes learning as a dynamic interaction among presage factors (student characteristics, prior knowledge, motivational beliefs), process behaviors (strategies and approaches), and product outcomes (learning results). This model provides the structural architecture for the present study: demographic variables, AI literacy, and EVT beliefs constitute the presage layer; LLM reliance type and intensity form the process layer; and self-reported writing outcomes represent the product layer. Two properties of the 3P Model are particularly significant. First, \citet{biggs1989} characterizes the model as ``an interactive system in equilibrium,'' where ``variations to any one component affect the whole system'' (p.~12). This mutual influence among presage, process, and product underpins the qualitative finding that students describe reliance trajectories rather than stable reliance states, reporting movement from Dependent toward Strategic reliance as AI literacy and EVT recalibrate through experience. Second, the model elucidates the measurement artifact observed when outcome instruments are structured around AI-attributed attainment rather than independent writing quality; such instruments capture the process layer (AI engagement) rather than the product layer (learning), thereby systematically misrepresenting the gap between the two.

\subsection{The Four-Type Taxonomy: Derived from the Integrated Architecture}

Building on this two-tier architecture and informed by writing process theory, which conceptualizes composing as a set of goal-directed cognitive processes \citep[p.~366]{flower1981} and distinguishes between ``knowledge telling'' and ``knowledge transforming'' strategies \citep[pp.~5--6]{bereiter1987}, as well as preliminary pilot data, four reliance types were identified. Each type is defined by a characteristic relationship between the student and AI's cognitive role in the writing process. These types vary along two theoretically grounded dimensions: the degree of \emph{authorial control} retained over ideational content, which is central to knowledge transforming (where the writer reprocesses their own knowledge rather than merely retrieving it; \citealp[p.~11]{bereiter1987}; cf.\ \citealp[p.~377]{flower1981}), and the extent of \emph{critical evaluation} applied to AI-generated outputs, which extends the writer's own ``evaluating and revising'' processes (\citealp[p.~374]{flower1981}; \citealp[p.~11]{bereiter1987}). Table~\ref{tab:blueprint} presents the complete typology, including defining characteristics, the cognitive role of AI, and theoretical grounding. Notably, prior frequency-based frameworks are unable to capture the distinction that Strategic and Dialogic reliance can involve substantial AI engagement while maintaining or even enhancing cognitive agency, whereas Dependent reliance involves similar surface-level engagement but systematically displaces cognitive agency.

\section{Literature Review}

\subsection{LLMs in Academic Writing: Adoption, Benefits, and Critical Tensions}

The integration of large language models (LLMs) into undergraduate academic writing has been both rapid and structurally disruptive, yet remains empirically under-theorized. Since the public release of ChatGPT in November 2022, these tools have evolved from novelty to essential infrastructure within three academic years. A 2025 survey of 1,041 UK undergraduates reported that 92\% now use AI in some form, an increase from 66\% a year earlier, and 88\% have used it specifically for assessments \citep{freeman2025}. Institutional policy and assessment frameworks, however, have not kept pace with this widespread adoption \citep{cotton2024}. Although research has documented adoption statistics, it has not sufficiently addressed a more fundamental question: what does it mean, in qualitatively distinct terms, to rely on LLMs during academic writing? Students utilize these tools at every stage of composition for a variety of purposes and intensities. In the pre-writing phase, LLMs support ideation and the formulation of research questions, addressing the persistent challenge of writer's block \citep{gilburt2024}. During drafting, students employ LLMs for paragraph generation, argument development, and structural organization. At the editing stage, grammar correction and stylistic polishing represent the least cognitively demanding applications \citep{cotton2024}.

The literature consistently identifies three primary benefits of LLMs: enhanced efficiency, improved accessibility, and personalized support. \citet{noy2023} demonstrated that access to ChatGPT reduced completion time for mid-level professional writing tasks by approximately 37\% while also increasing average output quality. For multilingual writers and students with writing-related disabilities, LLMs offer language support that exceeds traditional rule-based error detection, thereby lowering barriers to producing academic-standard prose. However, these advantages introduce a structural concern regarding homogenization. As many writers depend on a limited set of shared models, individual improvements in fluency may collectively result in a narrowing of expressive diversity, a phenomenon described as algorithmic monoculture \citep{kleinberg2021}. Empirical evidence substantiates this concern: in three preregistered studies involving 2,200 college-admissions essays, human writing expanded the collective semantic diversity of a corpus two to eight times more than GPT-4 writing, even after prompt and parameter adjustments \citep{moon2025}. Similarly, a controlled experiment with argumentative essays based on New York Times student-opinion prompts found that composing with an instruction-tuned model significantly reduced content diversity compared to unaided writing \citep{padmakumar2024}. These findings suggest that LLM assistance may shift academic prose toward statistically common rather than rhetorically distinctive expression, a risk that is particularly significant in undergraduate writing courses where cultivating an individual scholarly voice is a central educational objective.

These benefits are counterbalanced by documented cognitive costs. The field faces a performance paradox: while AI assistance enhances immediate task outcomes, it can also induce process-level disengagement that undermines intended learning objectives. In a randomized experiment comparing four conditions, ChatGPT, a human expert, writing analytics checklists, and no support ($N = 117$), the ChatGPT group achieved the highest essay scores but did not demonstrate corresponding gains in knowledge acquisition or transfer. This group also engaged in significantly fewer metacognitive self-regulation processes, particularly orientation and evaluation, than the human-expert and checklist groups, a phenomenon \citet{fan2025} refer to as ``metacognitive laziness.'' Additional evidence emerges at both the population and neural levels. In a mixed-methods study of 666 UK adults, including students, specialists, and managers, \citet{gerlich2025} identified a strong negative correlation between AI-tool use and critical-thinking performance ($r = -.68$), with cognitive offloading partially mediating this relationship (indirect effect $b = -.25$, $p < .001$). The author notes that the cross-sectional, self-report design supports correlation rather than causation. At the neural level, a preprint EEG study \citep{kosmyna2025} assigned 54 participants to LLM-assisted, search-engine-assisted, and unassisted writing groups. LLM users exhibited the weakest and least distributed brain connectivity; in early sessions, 83\% could not recall a sentence from an essay they had just produced, a pattern described as accumulating ``cognitive debt.'' This study also found that LLM-generated essays were statistically homogeneous within topics, reinforcing concerns about reduced diversity and supporting the argument that AI assistance alters both the writing process and the nature of its outputs. Notably, these studies compare users with non-users or different tools, but lack a framework for distinguishing qualitatively different forms of use. The present study directly addresses this gap.

\subsection{Writing Development Theory: Why Reliance Type Matters More Than Frequency}

A comprehensive understanding of why the type, rather than the frequency, of LLM reliance is educationally significant necessitates a grounding in foundational writing development theory.

\citeauthor{flower1981}'s \citeyearpar{flower1981} cognitive process model reconceptualized writing as a recursive, goal-directed problem-solving activity rather than a linear transcription of pre-formed ideas (pp.~365--366). Planning, translating, and reviewing are not simply procedural steps; they constitute the cognitive operations through which writers develop and refine their thinking. When students delegate these operations to LLMs, the resulting written product lacks the cognitive engagement that the model identifies as essential.

\citeauthor{bereiter1987}'s \citeyearpar{bereiter1987} distinction between knowledge-telling and knowledge-transforming writing highlights the most educationally significant dimension of reliance on LLMs. Knowledge-transforming entails a bidirectional, dialectical interaction between the problem space of content and the problem space of rhetoric, wherein writers use the act of writing to discover, test, and refine their ideas (Ch.~1). LLM assistance introduces fundamentally different risks depending on its application. When students use AI for knowledge-telling tasks, such as organizing already understood content or polishing grammar, the cognitive impact remains relatively limited because knowledge-telling requires minimal cognitive restructuring. In contrast, when LLMs are used for knowledge-transforming tasks, generating arguments, synthesizing sources, or developing analytical frameworks, students bypass the recursive problem-solving processes essential for both the written product and the writer's understanding. Dependent reliance, as defined in this study, aligns with this distinction: students who outsource ideation to LLMs may produce polished knowledge-telling texts while systematically under-developing the knowledge-transforming capacities required for advanced academic work.

\citeauthor{graham1994}'s \citeyearpar{graham1994,graham2000} Self-Regulated Strategy Development (SRSD) framework offers a developmental explanation for the importance of self-regulation in writing: expert writing relies on strategically deployed, self-regulatory processes that must be practiced until they become automatic, thereby freeing cognitive resources for higher-order concerns \citep[pp.~3--12]{graham2000}. When LLMs replace planning, monitoring, and evaluation, they may inhibit the accumulation of practice necessary for students to develop strategic writing capacity. The experimental evidence for metacognitive laziness reported by \citet{fan2025} aligns with this framework's prediction that external scaffolding can undermine rather than support the development of internal strategies.

The digital divide now extends beyond hardware and connectivity to include AI literacy as a key dimension of educational stratification. \citet{beckman2025} found that differences in students' AI literacy correspond to established patterns of digital inequality in access, capability, and outcomes, cautioning that emerging technologies tend to ``perpetuate and extend to existing digital divides'' (p.~7). This perspective builds on foundational definitions of AI literacy: \citet{long2020} describe it as ``a set of competencies that enables individuals to critically evaluate AI technologies; communicate and collaborate effectively with AI; and use AI as a tool'' (p.~2). \citet{ng2021} subsequently organized these competencies into four dimensions adapted from Bloom's taxonomy: ``know and understand, use and apply, evaluate and create, and ethical issues'' (p.~2). \citet{allen2024} proposed a distinct six-component framework for educational contexts, ED-AI Lit, encompassing ``Knowledge, Evaluation, Collaboration, Contextualization, Autonomy, and Ethics'' (p.~3). Across these frameworks, AI literacy determines not whether students use LLMs, but how they engage with them. Students with stronger foundational knowledge use these tools strategically, whereas those with weaker preparation are more susceptible to uncritical acceptance. A systematic review by \citet{zhai2024} found that users with lower subject-matter expertise are ``particularly prone to trust the AI's advice, even when it is incorrect'' (p.~27), leading to the incorporation of errors and biases that undermine academic quality. \citet{beckman2025} empirically illustrated this gap by identifying three student profiles differentiated by digital and AI literacy: lower-literacy ``novice'' and ``cautious'' users who were ``largely uncertain (and at times fearful)'' about the technology (p.~7), and higher-literacy ``enthusiastic'' users who used GenAI to enhance their learning (p.~5). This AI literacy gap intersects with socioeconomic stratification, compounding existing inequities.

\citet{warschauer2023} identify a critical dynamic in which LLMs offer language scaffolding that benefits multilingual writers, but writing competence is necessary to leverage this support effectively. They note that ``it takes a certain amount of privilege to fully exploit AI-generated writing'' (p.~3), and caution that the technology ``runs the risk of becoming yet another contributor to the same inequality that it has the potential to address'' (p.~3). The central paradox is developmental: higher-proficiency learners engage selectively with AI feedback, while lower-proficiency students are more likely to trust and rely on it uncritically (p.~3). As a result, ``the better students can write without AI, the better they will be prepared to write with it'' (p.~3). Students lacking foundational skills risk producing superficially polished texts that mask, rather than develop, their underlying capabilities. In Minority-Serving Institutions (MSIs) characterized by heterogeneous preparation, this paradox is particularly salient. The present study's finding that strategic reliance was predicted by higher prior AI literacy empirically confirms this mechanism.

This vulnerability is exacerbated by an institutional preparedness gap. In a late-2025 survey of 1,057 U.S. faculty, \citet{watson2026} found that 68\% reported their institutions had not prepared faculty to use GenAI for effective teaching and mentoring (p.~3), and 81\% believed GenAI would widen digital inequities (p.~4). The authors note that the sample was non-scientific and not generalizable (p.~5). Without systematic AI literacy development, students with the most prior preparation are best positioned to use LLMs productively, thereby deepening the inequities that Minority-Serving Institutions are intended to address.

The integration of LLMs has destabilized academic integrity frameworks that rely on source-matching detection. Because LLMs generate novel text that does not exist in any retrievable source, conventional plagiarism detection methods are rendered technically obsolete, and purpose-built replacements have proven similarly ineffective. In a systematic evaluation of fourteen AI-text detectors, \citet{weberwulff2023} concluded that ``the available detection tools are neither accurate nor reliable'' (Article~26), consistently misclassifying AI-generated text as human-written. These tools also produce higher false-positive rates among non-native English speakers; \citet{liang2023} found that detectors falsely flagged L2 writers' texts as AI-generated more than half the time, raising equity concerns as multilingual students face disproportionate scrutiny. \citet{eaton2023,eaton2025} describes the current context as a ``postplagiarism era'' in which ``hybrid human-AI writing will become the norm'' \citep[Article~23]{eaton2023}, shifting the integrity challenge from detecting copied text to teaching students to engage ethically with AI. This represents a transition from a policing model to an instructional one, where writers may delegate control over writing but retain responsibility for the content. \citet{kofinas2025} found that students often delegated to GenAI because they perceived its output ``as superior to the prose they could write themselves'' (p.~2), with delegation motivated not only by efficiency but also by self-doubt regarding academic capability. This finding aligns with the dynamics of self-efficacy in dependent reliance documented in the present study. Consequently, the most sustainable institutional response is not technical surveillance, but the cultivation of principled academic authorship through explicit instruction and AI literacy curricula.

The present study synthesizes three frameworks that address distinct yet complementary aspects of LLM reliance. The AI Literacy Framework \citep{long2020,ng2021,allen2024} accounts for why AI literacy predicts the \emph{type} of reliance students adopt: higher literacy encourages critical and evaluative engagement rather than mere suppression of use. Expectancy-Value Theory \citep{wigfield2000} explains why motivational beliefs influence the \emph{intensity} of reliance; the theory posits that achievement behavior is a function of ``expectancies for success, and the components of subjective task values'' (p.~68). Thus, students who believe they can use AI effectively (expectancy) and who place a high value on AI assistance (task value) are more likely to engage broadly and persistently. \citeauthor{biggs1993a}'s \citeyearpar{biggs1993a} Presage-Process-Product model offers a structural framework that links individual characteristics (presage) to engagement strategies (process) and outcomes (product), emphasizing that deep versus surface approaches are contextually variable rather than fixed traits. Integrating these frameworks yields a two-tiered predictive architecture: AI literacy determines the direction of engagement, while Expectancy-Value Theory determines its intensity. These are distinct systems that require separate interventions, a claim not previously operationalized within a single study design.

The reviewed literature reveals three convergent gaps that motivate the present study. Conceptually, existing frameworks conceptualize LLM engagement primarily as a frequency-based construct rather than a typological one, resulting in outcome measures that capture AI throughput rather than writing quality and systematically misclassify students demonstrating the most cognitive agency as underperformers. Institutionally, the literature is predominantly non-American and non-MSI, leaving the equity dynamics of AI literacy and access underexplored in contexts where preparation heterogeneity and first-generation enrollment are central rather than peripheral. Methodologically, large-scale surveys establish prevalence but do not explain underlying mechanisms, while small qualitative studies provide insight into meaning-making but lack generalizability. No prior study has employed a sequential explanatory mixed-methods design to measure reliance patterns at scale and then systematically investigate the meaning-making and motivational dynamics underlying those patterns within the same population. The present study addresses all three gaps.

\section{Methodology}

\subsection{Research Design and Rationale}

A sequential explanatory mixed-methods design was utilized \citep{creswell2018}. In this approach, a quantitative survey phase (Phase 1) addressed the primary research questions and informed the purposive sampling of qualitative participants for Phase 2. The design was based on a pragmatist epistemological stance, which prioritized research questions over predetermined philosophical commitments and treated quantitative and qualitative evidence as complementary tools selected for their appropriateness to the research problem \citep{morgan2014}. A purely quantitative design could establish prevalence and test hypothesized associations but could not explain the underlying reasons for students' reliance patterns or their ethical reasoning. In contrast, a purely qualitative design could reveal meaning-making processes but lacked the statistical power to test the theorized predictor structure in a representative sample. The sequential explanatory architecture addressed both limitations: Phase 1 established large-scale patterns, while Phase 2 provided in-depth explanations.

The qualitative phase incorporated three complementary data strands to triangulate the typology using multiple sources of evidence. The primary strand consisted of 14 semi-structured interviews, purposively sampled to maximize variation in reliance type, AI literacy level, first-generation status, and discipline. The second strand comprised open-ended responses from 35 participants who opted into a dedicated qualitative instrument embedded in the Phase 1 survey (hereafter SUR35). The third strand included a brief open-ended item completed by 361 of 382 survey respondents (94.5\%; hereafter SUR361), representing the largest qualitative source and enabling population-level qualitative analysis not possible through purposive subsampling alone. Together, these three strands yielded 1,435 coded instances across 47 codes and seven themes.

\subsection{Participants and Setting}

Participants were undergraduate students at a public R1 minority-serving institution (MSI) and Asian American and Native American Pacific Islander-Serving Institution (AANAPISI) in the mid-Atlantic, where over half of undergraduates identify with one or more minoritized racial or ethnic groups \citep{oir2025}. This institution was purposively selected because MSIs concentrate conditions of preparation heterogeneity, high first-generation enrollment, and broad reliance on federal financial aid, under which differential AI reliance is theorized to have the greatest equity consequences.

Recruitment was conducted through three channels over a six-week period: (1) email invitations distributed via departmental listservs coordinated by faculty liaisons across more than 30 academic departments representing STEM, social sciences, humanities, and professional programs; (2) in-class announcements delivered by cooperating instructors to supplement email outreach, particularly in large-enrollment courses; and (3) a university-wide undergraduate email list. To maximize participation rates, 50 randomly selected respondents received \$10 in monetary compensation, and all 14 interview participants received \$25. A staged reminder protocol targeting non-respondents was implemented throughout the collection period. Recruitment materials emphasized voluntary participation, confidentiality, and institutional IRB approval. After systematic data screening, including the removal of one Qualtrics label row, two preview submissions, and one case with more than 20\% item non-response, the final analytic sample comprised $N = 382$ participants.

Sample size was justified through a priori power analysis using G*Power 3.1.9.7 \citep{faul2007}. For hierarchical regression with eight predictors, $n = 382$ exceeded the minimum of $n = 236$ required to detect small-to-medium effects ($f^2 = 0.10$) at power $= .95$. For one-way ANOVA comparing four reliance types, $n = 382$ substantially exceeded the $n = 280$ needed for power $= .95$ at medium effect sizes. The moderation analysis target was $n = 395$ for small interaction effects ($f^2 = 0.02$); the achieved $n = 382$ was slightly below this threshold, indicating that non-significant moderation findings should be interpreted as inconclusive rather than as evidence of true null effects.

\subsection{Measures}

\subsubsection{Instrumentation}

\paragraph{Construct Definition} The LLM Reliance construct is operationally defined as the characteristic manner in which a writer engages large language models during academic writing, distinguished by two dimensions: the \emph{locus of cognitive control} (the extent to which the student retains authorial agency over ideational content) and the \emph{degree of critical evaluation} applied to AI-generated outputs prior to incorporation. This definition distinguishes reliance \emph{type}, a qualitative orientation, from reliance \emph{intensity}, which refers to the breadth and frequency of engagement and is governed by distinct predictor systems requiring separate measurement. Four dimensions were specified based on writing process theory \citep{bereiter1987,flower1981} and preliminary pilot data: Strategic, Instrumental, Dialogic, and Dependent. Full conceptual definitions and theoretical grounding for each dimension are provided in the theoretical framework section and Table~\ref{tab:blueprint}.

\begin{table*}
\caption{Construct Blueprint: LLM Reliance Scale and Companion Measures}
\label{tab:blueprint}
\begin{tabular}{llp{7.2cm}cl}
\toprule
\textbf{Construct} & \textbf{Dimension} & \textbf{Conceptual indicator} & \textbf{Final items} & \textbf{Source} \\
\midrule
\emph{LLM Reliance} & Strategic & Deliberate, goal-directed engagement; active verification and critical evaluation of AI outputs; full authorial control maintained throughout & 8 & New \\
 & Instrumental & Bounded use for mechanical tasks (outlining, sentence revision, summarizing) without delegation of ideational content & 4 & New \\
 & Dialogic & Iterative co-construction with AI as thinking partner; authorial agency preserved throughout the exchange & 4 & New \\
 & Dependent & Cognitive abdication; wholesale outsourcing of writing tasks with minimal critical oversight & 4 & New \\
\emph{AI Literacy} & Core knowledge & Understanding of LLM functioning, output evaluation, error detection, citation knowledge, and ethical self-regulation & 6 & Allen \& Kendeou (2024) \\
 & Prior Exposure & Breadth and depth of prior LLM experience across academic contexts & 7 & New \\
\bottomrule
\multicolumn{5}{p{16.5cm}}{\footnotesize\emph{Note.} All items used 7-point Likert scales (1 = \emph{Strongly Disagree}, 7 = \emph{Strongly Agree}). ``New'' = items developed specifically for this study against the construct definitions above. Item counts reflect the final retained instrument after pilot testing ($n = 143$). CE = Critical Evaluation sub-facet; DEL = Deliberate Use sub-facet of the Strategic subscale.} \\
\end{tabular}
\end{table*}

\paragraph{Item Generation} Items were generated from two sources: deductive sampling against the construct blueprint, with items anchored to each dimension's theoretical definition, and inductive refinement through preliminary pilot interview data in which participants described their LLM engagement practices. Strategic reliance items were adapted from \citeauthor{schraw1994}'s \citeyearpar{schraw1994} metacognitive awareness scales and \citeauthor{long2020}'s \citeyearpar{long2020} AI literacy framework. Dependent reliance items were adapted from \citeauthor{bandura1997}'s \citeyearpar{bandura1997} self-efficacy scale. Dialogic reliance items were informed by \citeauthor{vygotsky1978}'s \citeyearpar{vygotsky1978} sociocultural theory and collaborative learning frameworks. Instrumental reliance items were developed specifically for this study based on the blueprint. All items were drafted as first-person behavioral statements rated on a 7-point Likert scale (1 = Strongly Disagree, 7 = Strongly Agree) and reviewed for behavioral specificity, temporal specificity, and construct fidelity. Each item was designed to describe an observable action, anchored to current academic writing practice, and targeted to exactly one dimension of one construct \citep{shultz2014}.

\paragraph{Expert Content Review} Before pilot administration, item content was reviewed by a small group of colleagues with expertise in educational measurement and AI in education, who provided qualitative feedback on item clarity, construct coverage, and wording precision. This feedback informed revisions to five items identified in the pilot as producing inconsistent response patterns. Formal computation of item-level content validity indices (I-CVIs) was not conducted, a limitation of the current validation sequence. Future refinement of the LLM Reliance Scale should include a structured expert panel using the \citet{lynn1986} I-CVI procedure with a minimum of five raters and a retention threshold of I-CVI $\geq .78$.

\paragraph{Instrument Development} The LLM Reliance Scale and its companion measures were developed through a structured process: (1) construct definition and domain specification based on a theoretical blueprint, (2) item generation drawing on pilot interviews and existing frameworks, (3) content review, (4) pilot administration with item analysis ($n = 143$), and (5) main-study psychometric evaluation. Pilot administration led to three critical refinements: the number of open-ended survey questions was reduced from six to three due to low completion rates (38\% compared to 87\% for Likert items); wording was clarified on five items where pilot respondents showed inconsistent response patterns; and interview protocol sequencing was validated through three pilot interviews. Pilot internal consistency was acceptable across all subscales ($\alpha = .79$--$.88$). Table~\ref{tab:blueprint} presents the construct blueprint that guided item development, and the psychometric properties of the final main-study instrument are reported in Table~\ref{tab:reliability} and the Psychometric Evaluation section.

\subsubsection{LLM Reliance Scale}

The LLM Reliance Scale comprised 20 items across four subscales. The Strategic subscale (8 items) integrated two sub-facets: Deliberate Use (4 items measuring goal-directed, purposeful engagement) and Critical Evaluation (4 items measuring active oversight and verification of AI-generated outputs). The Instrumental subscale (4 items) assessed bounded, task-directed AI use for specific mechanical functions, such as outlining, sentence revision, and summarizing, without delegation of ideational content. The Dialogic subscale (4 items) captured iterative co-construction in which AI functions as a conversational thinking partner while the student retains authorial agency. The Dependent subscale (4 items) measured cognitive abdication: wholesale outsourcing of writing tasks with minimal critical oversight. All items used 7-point Likert scales (1 = Strongly Disagree, 7 = Strongly Agree).

\begin{table*}
\caption{Subscale Structure, Reliability, and Validity Estimates (Main Study, $N = 375$--$382$)}
\label{tab:reliability}
\begin{tabular}{lccccp{7.4cm}}
\toprule
\textbf{Subscale} & \textbf{k} & $\boldsymbol{\alpha}$ & $\boldsymbol{\omega}$ & \textbf{AVE} & \textbf{Representative item} \\
\midrule
\multicolumn{6}{l}{\emph{Deliberate, regulated engagement}} \\
Strategic & 8 & .786 & .789 & .331 & ``I verify facts in AI-generated text against authoritative sources.'' \\
Critical Evaluation facet & (above) & --- & --- & --- & ``I evaluate the logical consistency of AI output before incorporating it.'' \\
\multicolumn{6}{l}{\emph{Task-bounded engagement}} \\
Instrumental & 4 & .877 & .878 & .642 & ``I use generative AI tools to rephrase complex sentences in my drafts.'' \\
\multicolumn{6}{l}{\emph{Iterative co-construction}} \\
Dialogic & 4 & .859 & .860 & .607 & ``I engage in multiple prompt rounds to co-develop text with AI.'' \\
\multicolumn{6}{l}{\emph{Uncritical delegation}} \\
Dependent & 4 & .879 & .880 & .647 & ``I incorporate AI suggestions into drafts with minimal revision.'' \\
\multicolumn{6}{l}{\emph{Predictor scales}} \\
AI Literacy---Core & 6 & .775 & .795 & .402\textsuperscript{a} & ``I can detect errors and bias in AI-generated content.'' \\
AI Literacy---Exposure & 7 & .664\textsuperscript{b} & .653\textsuperscript{b} & .240\textsuperscript{a} & ``I have used generative AI for discipline-specific writing tasks.'' \\
Expectancy-Value & 5 & .923 & .924 & .709 & ``I am confident AI tools will enhance my academic performance.'' \\
\multicolumn{6}{l}{\emph{Writing process scales}} \\
Planning & 3 & .894 & .895 & .741 & ``I use AI to develop detailed outlines before drafting.'' \\
Drafting & 3 & .869 & .875 & .701 & ``I use AI to overcome writer's block more quickly.'' \\
Revising & 3 & .938 & .939 & .837 & ``I use AI to revise my draft more thoroughly.'' \\
Editing & 3 & .900 & .901 & .752 & ``I use AI to catch grammar and style errors.'' \\
\bottomrule
\multicolumn{6}{p{16.5cm}}{\footnotesize\emph{Note.} k = number of items; $\alpha$ = Cronbach's alpha (main study); $\omega$ = McDonald's omega; AVE = average variance extracted from a single-factor solution. All items used 7-point Likert scales (1 = Strongly Disagree, 7 = Strongly Agree). The Critical Evaluation facet and Deliberate Use facet together constitute the 8-item Strategic subscale; both representative items are listed to illustrate the two sub-facets. Writing-process items are shown without the shared preamble ``I use generative AI tools to\ldots'' for brevity. \textsuperscript{a} AVE below the conventional $\geq .50$ threshold. \textsuperscript{b} $\alpha$ and $\omega$ below the conventional $\geq .70$ threshold; acknowledged as a scale limitation.} \\
\end{tabular}
\end{table*}

\subsubsection{Predictor and Outcome Measures}

AI literacy was measured with 13 items across two subscales. Core AI Literacy (6 items adapted from \citealp{allen2024}) measured understanding of LLM processing, output evaluation, error detection, citation knowledge, and ethical self-regulation. Prior Exposure (7 items) assessed the breadth and depth of prior LLM experience across academic contexts. Expectancy-Value beliefs were measured with 5 items adapted from \citet{wigfield2000}, covering expectancy for success, efficiency value, intrinsic enjoyment, attainment value, and utility value. Writing process engagement was assessed with 12 items across four stages (Planning, Drafting, Revising, Editing; 3 items per stage), measuring AI-supported engagement at each stage. Writing outcomes were assessed using 6 single-item indicators (Quality, Self-Efficacy, Clarity, Grammar/Style, Originality, and Critical Thinking) adapted from \citet{braten2023}, \citet{bruning2013}, and \citet{tierney2002}, which capture perceived AI-mediated writing attainment.

\subsection{Scoring and Classification}

Subscale scores were computed as the mean of their constituent items, preserving the 1--7 metric so that scores remain interpretable relative to the scale midpoint (4.0) and are comparable across subscales of unequal length. A high subscale score indicates stronger endorsement of that reliance orientation; it does not indicate greater overall frequency of AI use. Reliance intensity, the continuous outcome in RQ2, was operationalized as the mean across all 20 reliance items.

Each participant was assigned a dominant reliance type corresponding to their highest subscale mean, provided that the mean reached a minimum endorsement threshold of 5.0 (above the scale midpoint), indicating a substantively adopted pattern rather than a marginal preference. This classification rule yielded the following: Strategic ($n = 131$, 34.3\%), Instrumental ($n = 118$, 30.9\%), Dialogic ($n = 116$, 30.4\%), and Dependent ($n = 17$, 4.5\%). All 382 participants met the threshold on at least one subscale, indicating that the abstention-mode population identified in the qualitative phase (13.0\% of SUR361 respondents reporting principled non-use) is not detectable through scale scores alone; they registered low-intensity strategic profiles quantitatively while articulating categorical refusal qualitatively. This is the basis for the Two-Model Architecture introduced in the Discussion.

\subsection{Psychometric Evaluation}

\subsubsection{Dimensionality: Exploratory Factor Analysis}

EFA was conducted on all 20 reliance items using principal-axis factoring with Promax rotation. Sampling adequacy was meritorious: KMO $= .926$, Bartlett's $\chi^2(190) = 4{,}539.14$, $p < .001$. Parallel analysis and the scree plot both supported retention of four factors, accounting for 57.2\% of common variance (eigenvalues: 8.10, 3.24, 1.25, 1.11). The pattern matrix (Table~\ref{tab:efa}) revealed three cleanly defined factors and one important cross-loading: F1 loaded all four Instrumental items (loadings .69--.84) and all four Dialogic items (loadings .59--.83); F2 loaded all four Dependent items (.69--.86); F3 loaded the four Critical Evaluation items of the Strategic subscale (.55--.88); and F4 loaded the four Deliberate Use items of the Strategic subscale (.45--.67). The Instrumental--Dialogic co-loading on F1 indicates that, at the indicator level, these two constructs share substantial empirical variance. This structural overlap is acknowledged as a scale limitation; the theoretical and qualitative distinction between them, Instrumental AI as a mechanical assistant versus Dialogic AI as a thinking partner, is supported by Theme 2 and Theme 6 of the qualitative analysis but requires indicator-level refinement in future scale development.

\begin{table}
\caption{EFA Pattern Matrix: Four-Factor Promax Solution ($N = 375$)}
\label{tab:efa}
\small
\setlength{\tabcolsep}{3pt}
\begin{tabular}{llcccc}
\toprule
\textbf{Item} & \textbf{Empirical factor} & \textbf{F1} & \textbf{F2} & \textbf{F3} & \textbf{F4} \\
\midrule
I1 (Instrumental) & F1: Engagement & \textbf{+.69*} & --- & --- & --- \\
I2 & & \textbf{+.84*} & --- & --- & --- \\
I3 & & \textbf{+.83*} & --- & --- & --- \\
I4 & & \textbf{+.79*} & --- & --- & --- \\
DL1 (Dialogic) & F1: Engagement & \textbf{+.64*} & --- & --- & --- \\
DL2 & & \textbf{+.71*} & --- & --- & --- \\
DL3 & & \textbf{+.59*} & --- & --- & --- \\
DL4 & & \textbf{+.83*} & --- & --- & --- \\
DP1 (Dependent) & F2: Dependent & --- & \textbf{+.69*} & --- & --- \\
DP2 & & --- & \textbf{+.81*} & --- & --- \\
DP3 & & --- & \textbf{+.86*} & --- & --- \\
DP4 & & --- & \textbf{+.83*} & --- & --- \\
S5 (CE---Strategic) & F3: Crit.\ Eval. & --- & --- & \textbf{+.82*} & --- \\
S6 & & --- & --- & \textbf{+.88*} & --- \\
S7 & & --- & --- & \textbf{+.72*} & --- \\
S8 & & +.39 & --- & \textbf{+.55*} & --- \\
S1 (DEL---Strategic) & F4: Delib.\ Use & --- & --- & --- & \textbf{+.58*} \\
S2 & & --- & --- & --- & \textbf{+.67*} \\
S3 & & --- & --- & --- & \textbf{+.45*} \\
S4 & & --- & +.39 & --- & \textbf{+.48*} \\
\bottomrule
\multicolumn{6}{p{8.2cm}}{\footnotesize\emph{Note.} F1 = Engagement factor (Instrumental and Dialogic items co-load); F2 = Dependent; F3 = Critical Evaluation facet of Strategic; F4 = Deliberate Use facet of Strategic. Pattern coefficients below .30 are suppressed and replaced with an em dash. Bold type and an asterisk (*) indicate a primary loading ($\lambda \geq .40$). Unbolded coefficients without asterisks are cross-loadings retained for transparency. CE = Critical Evaluation sub-facet; DEL = Deliberate Use sub-facet of the Strategic subscale. The co-loading of Instrumental and Dialogic items on F1 is discussed as a scale limitation in the Psychometric Evaluation section.} \\
\end{tabular}
\end{table}

\subsubsection{Reliability}

Main-study internal consistency was indexed by both Cronbach's $\alpha$ and McDonald's $\omega$ (Table~\ref{tab:reliability}). Acceptable-to-excellent reliability was achieved for Instrumental ($\alpha = .877$, $\omega = .878$), Dependent ($\alpha = .879$, $\omega = .880$), Dialogic ($\alpha = .859$, $\omega = .860$), Expectancy-Value ($\alpha = .923$, $\omega = .924$), and all writing-process subscales ($\alpha = .869$--$.938$). The Strategic subscale produced lower reliability ($\alpha = .786$, $\omega = .789$), consistent with the EFA finding that its two facets (Deliberate Use and Critical Evaluation) load on separate factors; the 8-item composite indexes a broad orientation rather than a unidimensional latent trait. The AI Literacy, Exposure subscale fell below the conventional threshold ($\alpha = .664$, $\omega = .653$), reflecting heterogeneous item content spanning different types of AI experience; this subscale is used as a predictor of reliance intensity rather than as a stand-alone construct score, and its sub-threshold consistency is acknowledged as a limitation.

\subsubsection{Convergent and Discriminant Validity}

Convergent validity was supported by a theoretically predicted positive correlation between Strategic reliance and the AI Literacy composite ($r = .610$) and by average variance extracted (AVE) values that were acceptable for three of four reliance subscales (Instrumental $= .642$, Dialogic $= .607$, Dependent $= .647$). The Strategic subscale's AVE (.331) fell below the conventional .50 threshold, attributable to its two-facet structure identified in EFA; facet-level AVEs are acceptable.

Discriminant validity was evaluated through three converging criteria. First, the complete inter-factor correlation matrix (Table~\ref{tab:discriminant}, lower triangle) showed theoretically expected patterns: the Strategic--Dependent correlation was lowest ($r = .187$), consistent with their conceptual opposition as restrained versus wholesale AI engagement, while Instrumental, Dialogic, and Dependent intercorrelations were higher ($r = .530$--$.774$), reflecting their shared characteristic of high AI engagement volume. Second, the Fornell--Larcker criterion held for all pairs except marginally for Instrumental--Dialogic ($\sqrt{\mathrm{AVE}}_{\mathrm{Instr}} = .801$ vs.\ $r = .774$; $\sqrt{\mathrm{AVE}}_{\mathrm{Dial}} = .779$ vs.\ $r = .774$). Third, the heterotrait--monotrait ratio \citep[HTMT;][]{henseler2015} fell below the conservative .85 threshold for all pairs except Instrumental--Dialogic (HTMT $= .891$). These results confirm that Dependent, Strategic, and the other pairings are empirically distinct but reveal that Instrumental and Dialogic share a level of empirical variance that challenges their separation as independent constructs at the indicator level, an expected consequence of the EFA's F1 co-loading and a priority for future scale refinement.

\begin{table}
\caption{Factor Intercorrelations, $\sqrt{\mathrm{AVE}}$ (Diagonal), and HTMT Ratios for Reliance Subscales}
\label{tab:discriminant}
\small
\setlength{\tabcolsep}{4pt}
\begin{tabular}{lcccc}
\toprule
 & \textbf{1. Strat.} & \textbf{2. Instr.} & \textbf{3. Dial.} & \textbf{4. Dep.} \\
\midrule
1. Strategic & \textbf{.575} & .647 & .679 & .462 \\
2. Instrumental & .548 & \textbf{.801} & .891\textsuperscript{a} & .604 \\
3. Dialogic & .561 & .774 & \textbf{.779} & .638 \\
4. Dependent & .187 & .530 & .553 & \textbf{.804} \\
\bottomrule
\multicolumn{5}{p{8.2cm}}{\footnotesize\emph{Note.} Lower triangle = Pearson inter-factor correlations. Bold diagonal = square root of average variance extracted ($\sqrt{\mathrm{AVE}}$). Upper triangle = heterotrait--monotrait ratio of correlations (HTMT; Henseler et al.\ 2015). Discriminant validity is supported when (a) $\sqrt{\mathrm{AVE}}$ in each row exceeds the off-diagonal correlations in that row (Fornell--Larcker criterion) and (b) all HTMT ratios fall below the conservative threshold of .85. AVE = average variance extracted. HTMT = heterotrait--monotrait ratio. \textsuperscript{a} HTMT $= .891$ for the Instrumental--Dialogic pair, exceeding the .85 threshold; discriminant validity for this pair is not supported.} \\
\end{tabular}
\end{table}

\subsubsection{Criterion Validity and Common Method Variance}

Criterion validity was supported by significant one-way ANOVA effects of reliance type across all ten writing process and outcome variables (all $p < .001$; $\eta^2$ range $= .166$--$.329$), confirming that type membership systematically predicts theoretically expected behavioral differences. Known-groups validity was additionally supported by the multinomial logistic regression finding that AI literacy was the strongest predictor of type membership: each unit increase in AI literacy was associated with approximately a tenfold reduction in the odds of Dependent versus Strategic classification ($OR = 0.11$, $p < .001$), confirming that high-AI-literacy students disproportionately exhibit Strategic reliance.

Common method variance (CMV) was assessed using Harman's single-factor test, which produced a first unrotated factor accounting for 42.9\% of total variance, below the 50\% threshold, suggesting that common method variance was not a serious concern. Missing data were screened using \citeauthor{little1988}'s \citeyearpar{little1988} MCAR test, $\chi^2(16) = 18.43$, $p = .301$, confirming that data were missing completely at random; the overall missing rate was 0.26\%, and listwise deletion was employed for all multivariate analyses.

\paragraph{Fairness and Measurement Invariance} The equity argument this study advances, that AI reliance patterns carry differential consequences for first-generation, multilingual, and minoritized students at MSIs, presupposes that the LLM Reliance Scale measures the same constructs in the same metric across demographic groups. Measurement invariance across gender, first-generation status, and race/ethnicity was not tested in this study, representing a limitation of the current validation evidence. Future work should implement a configural--metric--scalar invariance sequence using multigroup CFA, with invariance inferred from $\Delta$CFI $\leq .010$ and $\Delta$RMSEA $\leq .015$ \citep{chen2007,putnick2016}. Scalar invariance is a necessary condition for the meaningful comparison of latent means across subgroups. Item-level differential item functioning (DIF) analysis should also be conducted to identify any items whose response patterns differ across groups in ways not explained by group differences on the underlying construct \citep{shultz2014}. Until invariance is established, between-group comparisons of reliance type prevalence and intensity should be treated as exploratory rather than confirmatory.

\paragraph{Intended Use and Interpretive Boundaries} Following best practice for applied educational instruments, the intended use of the LLM Reliance Scale is stated explicitly \citep{shultz2014}. The scale is designed to classify undergraduate writers according to qualitatively distinct patterns of AI-supported writing behavior for the purposes of research, AI literacy assessment, and educational intervention design.

\emph{Appropriate uses.} The instrument is appropriate for: (1) research examining the relationship between reliance patterns and learning outcomes; (2) AI literacy program evaluation, in which pre- and post-instruction reliance profiles assess whether instruction shifts students from uncritical to strategic engagement; and (3) educational intervention design, in which student reliance profiles inform differentiated instructional responses.

\emph{Inappropriate uses.} The instrument is expressly not designed for: (1) grading students or evaluating individual academic performance, as subscale scores reflect orientation toward AI engagement rather than writing quality or academic achievement; (2) academic misconduct detection, as Strategic users, those exercising the greatest cognitive restraint, produce the lowest scores by design, rendering low scores an unreliable indicator of problematic use; and (3) evaluating writing quality or cognitive ability, as the measurement artifact finding in this study demonstrates that AI-reliance outcome instruments capture AI throughput rather than independent writing competence.

\emph{Population boundaries.} The scale was developed and validated with undergraduate students at a public R1 minority-serving institution in the United States. Transfer to graduate students, K--12 populations, faculty, or professional contexts should be preceded by revalidation, as AI engagement dynamics and construct relevance may differ meaningfully across these groups. Transfer to non-English-language contexts requires translation validation, including back-translation, bilingual expert review, and independent psychometric evaluation in the target population.

\subsection{Phase 2 Qualitative Procedures}

Following Phase 1 data analysis, maximum variation sampling identified 14 interview participants representing diversity across reliance types, AI literacy levels, first-generation status, and discipline. Selected participants received personalized invitations describing the study purpose, estimated duration (approximately 60 minutes), and compensation (\$25). Interviews were conducted synchronously via Webex video conference, with video disabled to reduce self-consciousness, or asynchronously via structured written response for students facing scheduling constraints; pilot testing confirmed comparable data richness across modalities. Recordings were transcribed verbatim using Webex automated transcription, followed by manual verification of technical terminology. All participants received transcripts via email with a 72-hour member-checking review period \citep{lincoln1985}. The interview protocol comprised 13 core questions across four sections: LLM usage contexts, reliance-type self-perception, writing development impact, and ethical reasoning.

\subsection{Analytic Approach}

Quantitative analyses followed \citet{tabachnick2019}, with $\alpha = .05$ throughout and distributional assumptions (normality, homoscedasticity, multicollinearity via VIF) checked before each model. RQ1 was addressed through one-way ANOVA, Welch's F, where Levene's test indicated heterogeneous variances, with Tukey HSD or Games--Howell post-hoc comparisons and Cohen's $d$ effect sizes, complemented by multiple regression entering reliance intensity and dummy-coded reliance type as predictors of each outcome. RQ2 was addressed through four-block hierarchical multiple regression predicting reliance intensity (demographics $\rightarrow$ AI Literacy composite $\rightarrow$ AI Literacy--Exposure $\rightarrow$ Expectancy-Value), with each block's incremental contribution tested via $\Delta R^2$; reliance type membership was modeled through multinomial logistic regression with Strategic as the reference category. RQ3 was addressed using moderated multiple regression \citep{aiken1991}, with all continuous predictors mean-centered prior to interaction-term computation; significant interactions were probed using simple slopes analysis at $\pm 1$ SD of each moderator.

\subsubsection{Qualitative Analysis}

The three qualitative strands were analyzed using \citeauthor{braun2019}'s \citeyearpar{braun2019} reflexive thematic analysis in NVivo 15, proceeding through six recursive phases: familiarization, initial coding, theme development, theme review, theme definition, and reporting. Analysis was primarily inductive but theoretically informed by the three guiding frameworks. Rigor was established through four trustworthiness strategies: (1) researcher reflexivity through reflexive journaling maintained throughout all analytic phases; (2) thick description enabling transferability assessment; (3) member checking with four interview participants who confirmed the representational accuracy of their accounts; and (4) cross-strand corroboration, in which patterns were not reported as themes until confirmed independently across at least two of the three qualitative strands. Abstention-mode participants, principled non-users whose near-zero reliance scores reflect values-based categorical refusal rather than low AI literacy or low exposure, were identified through spontaneous, values-grounded statements of refusal in SUR361, not through low scale scores, which is what permits them to be distinguished analytically from low-intensity strategic restrainers.

\subsubsection{Integration}

Integration of the two phases followed a connecting-and-merging logic \citep{fetters2013}: quantitative classification structured the purposive sampling for Phase 2, and qualitative meta-inferences were developed by arraying the seven themes against the quantitative findings they explained, confirmed, or complicated. The measurement-artifact finding, that Strategic users score lowest on every outcome measure, emerged from this integration: the quantitative pattern was identified first, and qualitative evidence then provided the interpretive mechanism (Theme 2, Theme 5) explaining why outcome instruments capture AI throughput rather than writing quality. Researcher positionality, as writing instructors and AI education researchers at the study institution, was monitored through reflexive memoing and treated as an analytic resource informing rather than distorting interpretation.

\section{Results}

\subsection{Descriptive Statistics and Reliance Type Distribution}

Dominant reliance type was determined by the highest subscale mean score with a minimum threshold of 5.0 indicating meaningful pattern adoption. The resulting distribution was: Strategic ($n = 131$, 34.3\%), Instrumental ($n = 118$, 30.9\%), Dialogic ($n = 116$, 30.4\%), and Dependent ($n = 17$, 4.5\%). At the sample level, Strategic reliance recorded the highest mean ($M = 4.33$, $SD = 1.08$), reflecting that deliberate, regulated AI engagement was normative in this population, while Dependent reliance was notably low ($M = 2.64$, $SD = 1.50$), indicating that uncritical wholesale outsourcing was not. Among predictors, AI Literacy Composite scored at $M = 4.48$ ($SD = 0.91$) and Expectancy-Value beliefs at $M = 4.05$ ($SD = 1.68$), both near the scale midpoint and exhibiting substantial variance. Table~\ref{tab:descriptives} presents full descriptive statistics for all study variables.

\begin{table}
\caption{Descriptive Statistics for All Study Variables ($N = 379$--$382$).}
\label{tab:descriptives}
\small
\setlength{\tabcolsep}{3.4pt}
\begin{tabular}{lccccccc}
\toprule
\textbf{Variable} & \textbf{N} & \textbf{M} & \textbf{SD} & \textbf{Min} & \textbf{Max} & \textbf{Skew} & \textbf{Kurt} \\
\midrule
\multicolumn{8}{l}{\emph{LLM Reliance Types}} \\
Instrumental & 382 & 3.92 & 1.80 & 1.00 & 7.00 & $-$0.27 & $-$1.08 \\
Strategic & 382 & 4.33 & 1.08 & 1.00 & 7.00 & $-$0.32 & 0.40 \\
Dependent & 382 & 2.64 & 1.50 & 1.00 & 7.00 & 0.68 & $-$0.58 \\
Dialogic & 382 & 4.29 & 1.70 & 1.00 & 7.00 & $-$0.51 & $-$0.67 \\
\multicolumn{8}{l}{\emph{Individual Difference Predictors}} \\
AI Literacy--Exposure & 382 & 4.03 & 1.00 & 1.00 & 7.00 & 0.05 & $-$0.16 \\
AI Literacy--Core & 382 & 4.92 & 1.09 & 1.67 & 7.00 & $-$0.27 & $-$0.07 \\
AI Literacy Composite & 382 & 4.48 & 0.91 & 1.93 & 7.00 & $-$0.02 & $-$0.23 \\
Expectancy-Value & 382 & 4.05 & 1.68 & 1.00 & 7.00 & $-$0.35 & $-$0.89 \\
\multicolumn{8}{l}{\emph{Writing Process Engagement}} \\
Planning & 379 & 3.69 & 1.82 & 1.00 & 7.00 & $-$0.15 & $-$1.16 \\
Drafting & 380 & 3.51 & 1.78 & 1.00 & 7.00 & $-$0.03 & $-$1.15 \\
Revising & 382 & 3.96 & 1.97 & 1.00 & 7.00 & $-$0.33 & $-$1.27 \\
Editing & 382 & 4.18 & 1.99 & 1.00 & 7.00 & $-$0.39 & $-$1.16 \\
\multicolumn{8}{l}{\emph{Writing Outcomes}} \\
Quality & 380 & 4.02 & 2.09 & 1.00 & 7.00 & $-$0.27 & $-$1.38 \\
Self-Efficacy & 382 & 3.65 & 2.06 & 1.00 & 7.00 & 0.03 & $-$1.39 \\
Clarity & 381 & 3.93 & 2.05 & 1.00 & 7.00 & $-$0.27 & $-$1.39 \\
Grammar/Style & 379 & 4.39 & 2.28 & 1.00 & 7.00 & $-$0.41 & $-$1.41 \\
Originality & 380 & 3.03 & 1.86 & 1.00 & 7.00 & 0.37 & $-$1.20 \\
Critical Thinking & 380 & 3.47 & 1.95 & 1.00 & 7.00 & 0.08 & $-$1.27 \\
\bottomrule
\multicolumn{8}{p{8.2cm}}{\footnotesize\emph{Note.} All items measured on 7-point Likert scales (1 = Strongly Disagree, 7 = Strongly Agree). Scale midpoint = 4.0. Writing outcome variables are single-item measures.} \\
\end{tabular}
\end{table}

\subsection{RQ1: Reliance Type Differences in Writing Process Engagement and Outcomes}

One-way ANOVA produced statistically significant differences across all ten dependent variables (all $p < .001$) with large effect sizes throughout ($\eta^2$ range: .166--.329). The pattern was consistent and theoretically consequential: Strategic users reported the lowest engagement and outcome scores across every variable, while Dependent users reported the highest.

For Writing Process Engagement, Planning showed the most pronounced differentiation ($F = 55.21$, $p < .001$, $\eta^2 = .306$). Strategic users scored nearly two standard deviations below Instrumental users on Planning ($M = 2.32$ vs.\ $4.62$; $d = 1.49$, $p < .001$), with Dependent users reporting the most extensive AI-assisted planning ($M = 4.82$). This pattern replicated across Drafting ($\eta^2 = .324$), Revising ($\eta^2 = .297$), and Editing ($\eta^2 = .254$).

For Writing Outcomes, Quality ($F = 61.32$, $p < .001$, $\eta^2 = .329$) showed Strategic users at $M = 2.37$ ($SD = 1.77$) and Dependent users at $M = 5.24$ ($SD = 1.15$). Self-Efficacy ($F = 42.70$, $p < .001$, $\eta^2 = .253$) showed Strategic users at $M = 2.24$ ($SD = 1.71$) and Dependent users at $M = 5.12$ ($SD = 1.22$). Critical Thinking ($F = 34.61$, $p < .001$, $\eta^2 = .216$) showed Strategic users at $M = 2.28$ ($SD = 1.79$) and Dependent users at $M = 4.18$ ($SD = 1.47$). Originality produced the most theoretically significant pattern, with Strategic users at $M = 1.85$ ($SD = 1.44$), substantially below the scale midpoint, and Dependent users at $M = 4.35$ ($SD = 1.32$). Table~\ref{tab:anova} presents all group means, standard deviations, F statistics, and effect sizes.

\begin{table*}
\caption{Group Means, Standard Deviations, and ANOVA Results by Reliance Type ($N = 379$--$382$).}
\label{tab:anova}
\begin{tabular}{lcccccc}
\toprule
\textbf{Variable} & \makecell{\textbf{Instrumental}\\($n = 118$)\\M (SD)} & \makecell{\textbf{Strategic}\\($n = 131$)\\M (SD)} & \makecell{\textbf{Dependent}\\($n = 17$)\\M (SD)} & \makecell{\textbf{Dialogic}\\($n = 116$)\\M (SD)} & \textbf{F} & $\boldsymbol{\eta^2}$ \\
\midrule
\multicolumn{7}{l}{\emph{Writing Process Engagement}} \\
Planning & 4.62 (1.46) & 2.32 (1.61) & 4.82 (0.99) & 4.11 (1.55) & 55.21*** & 0.306 \\
Drafting & 4.31 (1.42) & 2.11 (1.39) & 4.47 (1.34) & 4.11 (1.62) & 60.12***\textsuperscript{\dag} & 0.324 \\
Revising & 4.76 (1.48) & 2.48 (1.78) & 4.84 (1.30) & 4.68 (1.72) & 53.24*** & 0.297 \\
Editing & 4.90 (1.49) & 2.79 (1.98) & 5.06 (1.14) & 4.88 (1.71) & 42.80***\textsuperscript{\dag} & 0.254 \\
\multicolumn{7}{l}{\emph{Writing Outcomes}} \\
Quality & 4.97 (1.64) & 2.37 (1.77) & 5.24 (1.15) & 4.73 (1.79) & 61.32*** & 0.329 \\
Self-Efficacy & 4.50 (1.77) & 2.24 (1.71) & 5.12 (1.22) & 4.15 (1.95) & 42.70***\textsuperscript{\dag} & 0.253 \\
Clarity & 4.91 (1.65) & 2.47 (1.87) & 4.62 (1.36) & 4.47 (1.77) & 46.74*** & 0.271 \\
Grammar/Style & 5.00 (1.86) & 3.12 (2.40) & 5.47 (1.50) & 5.05 (2.00) & 24.81***\textsuperscript{\dag} & 0.166 \\
Originality & 3.99 (1.76) & 1.85 (1.44) & 4.35 (1.32) & 3.21 (1.72) & 40.88***\textsuperscript{\dag} & 0.246 \\
Critical Thinking & 4.44 (1.61) & 2.28 (1.79) & 4.18 (1.47) & 3.73 (1.82) & 34.61*** & 0.216 \\
\bottomrule
\multicolumn{7}{p{15.5cm}}{\footnotesize\emph{Note.} All variables measured on 7-point scales (1 = Strongly Disagree, 7 = Strongly Agree). \textsuperscript{\dag} Welch's F reported; degrees of freedom adjusted. $\eta^2$ interpreted per Cohen [1988]: $\geq .14$ = large. *** $p < .001$. Group $n$s vary slightly across outcomes due to listwise deletion.} \\
\end{tabular}
\end{table*}

\begin{figure}
\centering
\includegraphics[width=\linewidth]{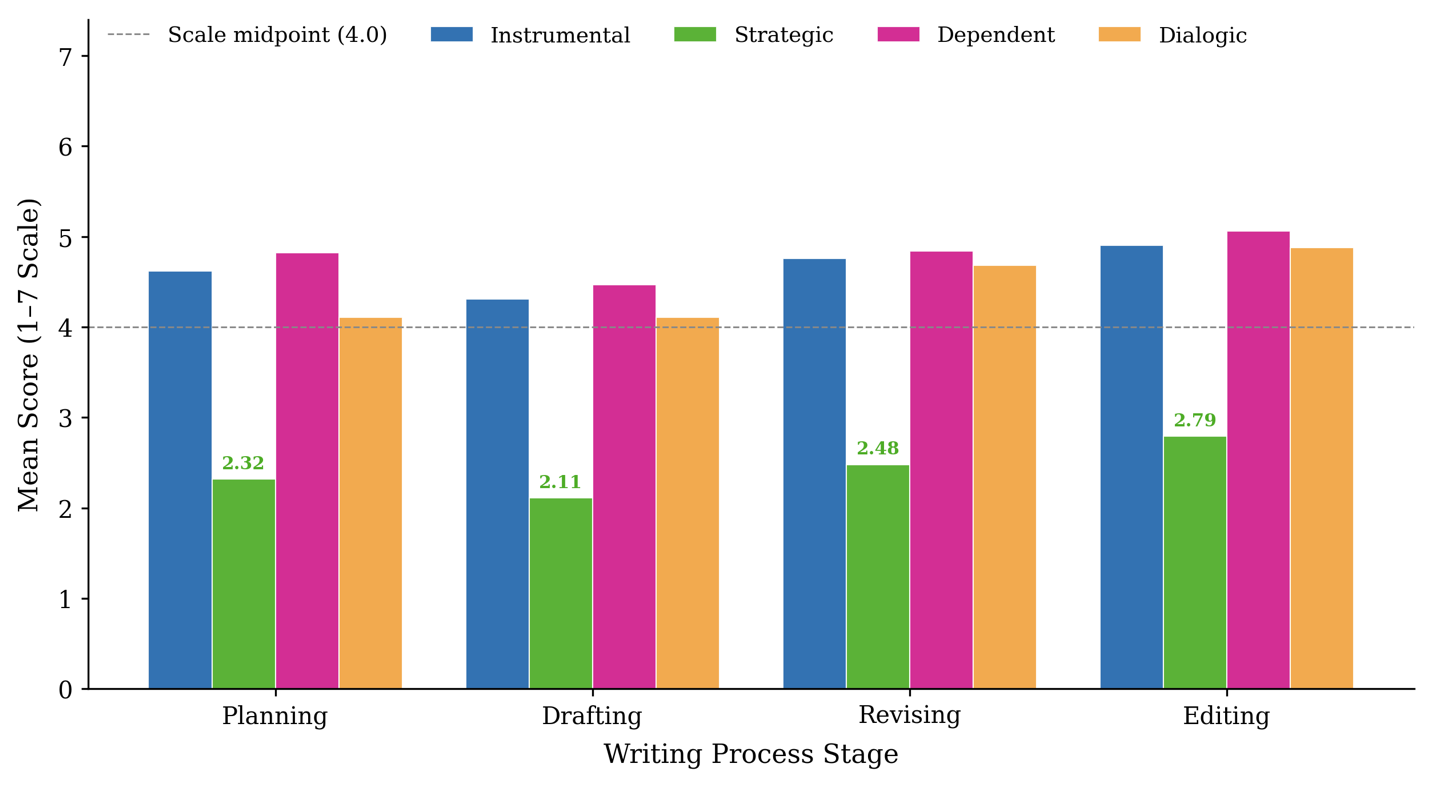}
\caption{Mean Writing Process Engagement by Dominant LLM Reliance Type (1 = Strongly Disagree, 7 = Strongly Agree; $N = 382$).}
\Description{Bar chart of mean writing process engagement scores at each of four writing stages, grouped by reliance type. Strategic users score lowest at every stage.}
\label{fig:process}
\end{figure}

\begin{figure}
\centering
\includegraphics[width=\linewidth]{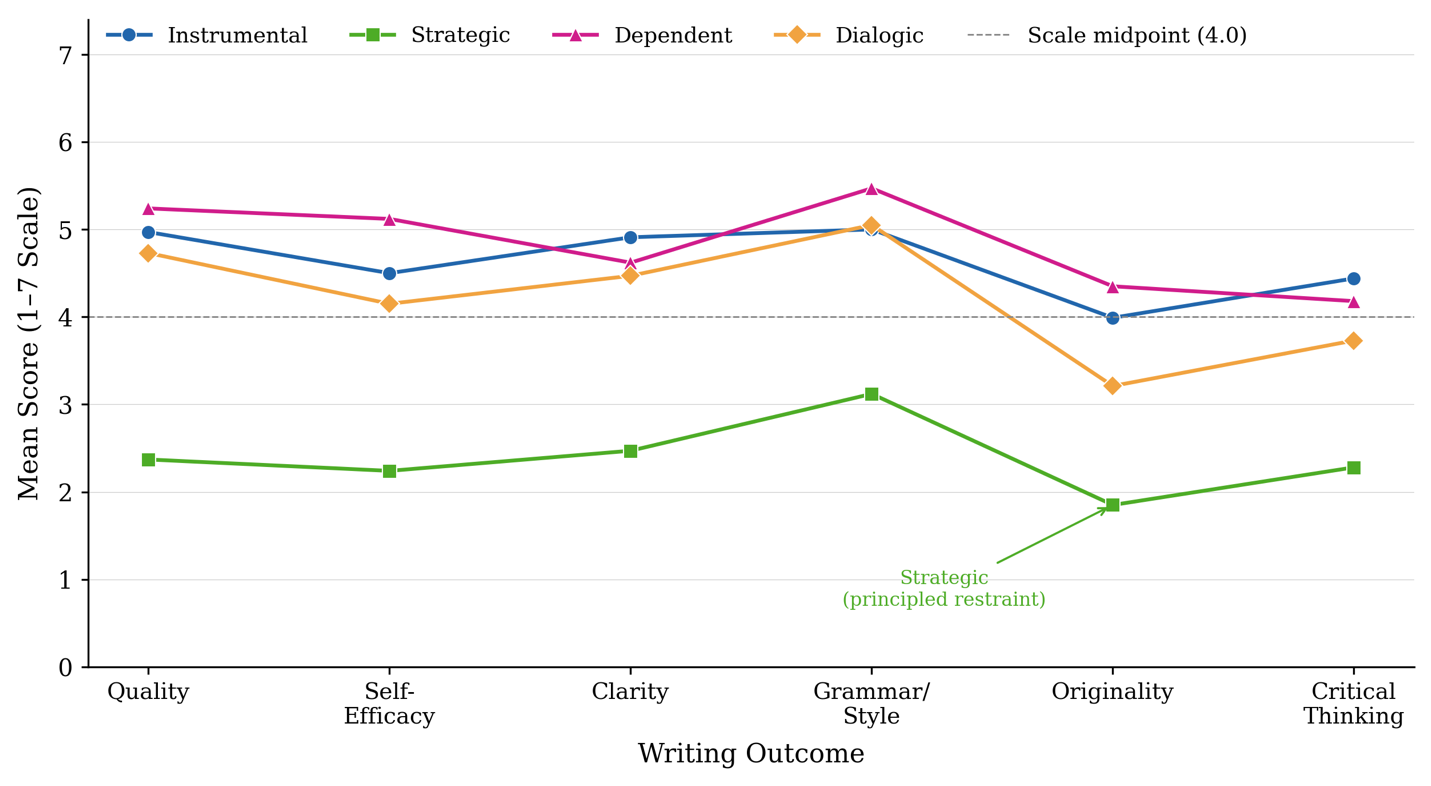}
\caption{Mean Writing Outcomes by Dominant LLM Reliance Type (1 = Strongly Disagree, 7 = Strongly Agree; $N = 382$).}
\Description{Line chart of mean writing outcome scores on six outcome measures, grouped by reliance type. Strategic users score lowest on every outcome.}
\label{fig:outcomes}
\end{figure}

Post-hoc pairwise comparisons confirmed that the Strategic--Dependent contrast produced the largest effect sizes in the study (Planning: $d = 1.61$; Drafting: $d = 1.76$; Quality: $d = 1.73$), establishing these as the most divergent behavioral profiles. Instrumental, Dependent, and Dialogic users did not significantly differ from one another on most variables, with notable exceptions for Originality (Instrumental $>$ Dialogic: $d = 0.45$; $p = .010$) and Critical Thinking (Instrumental $>$ Dialogic: $d = 0.41$; $p = .017$). Multiple regression confirmed that Reliance Intensity was the strongest and most consistent predictor across all outcome models ($\beta$ range: .598--.811, all $p < .001$; $R^2$ range: .403--.737), with the Strategic type dummy independently predicting lower AI-assisted attainment on Planning ($\beta = -.173$), Drafting ($\beta = -.192$), Quality ($\beta = -.256$), Self-Efficacy ($\beta = -.255$), and Originality ($\beta = -.259$) beyond overall intensity. Figures~\ref{fig:process} and~\ref{fig:outcomes} display these group means across the four writing-process stages and six writing-outcome measures, respectively.

\subsection{The Measurement Artifact: Interpreting Strategic Users' Suppressed Scores}

The pattern of Strategic users scoring lowest on every outcome variable requires direct interpretive intervention. A naive reading would suggest that strategic restraint underperforms relative to other reliance modes, a finding that would be theoretically incoherent given the construct's definition. The correct interpretation emerges from close examination of item wording. Every outcome item in this study, and in the published literature, employs the structure ``I use generative AI tools to [achieve X].'' These items capture perceived AI-mediated attainment, not independent writing quality or cognitive growth. Students who deliberately refrain from using AI for most writing functions will score low on these items by design. Strategic users' suppressed scores reflect their principled restraint, not their writing capacity.

Qualitative evidence confirmed this reinterpretation with precision. Tyler, a Mechanical Engineering junior representing the study's most pronounced strategic profile, described approaching AI verification as ``trying to disprove it almost. So, I'll go back to the textbook.'' His near-zero Originality score captured how little AI contributed to his writing, not how little originality his writing contained. This finding has field-wide implications: any study using analogous items to compare users and non-users, or frequent and infrequent users, is not measuring writing quality but AI throughput. The construct validity crisis this represents is not marginal. It is architectural.

\subsection{RQ2: Predictors of Reliance Intensity and Type Membership}

The four-block hierarchical regression produced a well-specified model with $R^2 = .722$ (adj.\ $R^2 = .716$). Block 1 demographics explained $R^2 = .061$ ($p < .001$), with First-Generation status ($\beta = .153$, $p = .008$) and STEM major ($\beta = .127$, $p = .013$) as the only individually significant predictors. Block 2 added AI Literacy composite ($\Delta R^2 = .264$, $F(1, 368) = 143.10$, $p < .001$; $\beta = .519$, $p < .001$), representing the largest single-block increment. Block 3 added AI Literacy-Exposure ($\Delta R^2 = .141$, $F(1, 367) = 96.76$, $p < .001$; $\beta = .755$, $p < .001$), with its entry reducing AI Literacy composite to non-significance ($\beta = -.143$, $p = .141$), indicating that prior exposure is the more proximal predictor of reliance breadth. Block 4 added Expectancy-Value beliefs, producing the strongest increment ($\Delta R^2 = .256$, $F(1, 366) = 335.92$, $p < .001$; $\beta = .630$, $p < .001$) and the largest unique contribution. In the full model, only AI Literacy-Exposure ($\beta = .308$, 95\% CI [0.189, 0.427]) and Expectancy-Value ($\beta = .637$, 95\% CI [0.568, 0.705]) retained significance. Table~\ref{tab:hierarchical} presents the full hierarchical regression model.

\begin{table}
\caption{Hierarchical Regression: Predictors of Overall LLM Reliance Intensity ($N = 375$).}
\label{tab:hierarchical}
\small
\setlength{\tabcolsep}{2pt}
\begin{tabular}{lcccccc}
\toprule
\textbf{Predictor} & \makecell{\textbf{Blk 1}\\$\beta$} & \makecell{\textbf{Blk 2}\\$\beta$} & \makecell{\textbf{Blk 3}\\$\beta$} & \makecell{\textbf{Blk 4}\\$\beta$} & \makecell{\textbf{p}\\(Blk 4)} & \textbf{VIF} \\
\midrule
\multicolumn{7}{l}{\emph{Block 1: Demographics}} \\
Gender (woman) & .049 & .051 & .032 & .029 & .311 & 1.11 \\
First-generation & .153** & .130** & .102* & .040 & .212 & 1.08 \\
STEM major & .127** & .089* & .111** & .027 & .348 & 1.10 \\
Lower SES & .092 & .099* & .064 & .048 & .126 & 1.11 \\
CGPA & .047 & .010 & $-$.006 & $-$.013 & .660 & 1.05 \\
\multicolumn{7}{l}{\emph{Block 2: AI Literacy}} \\
AI Literacy composite & --- & .519*** & $-$.143 & $-$.028 & .506 & 1.30 \\
\multicolumn{7}{l}{\emph{Block 3: AI Exposure}} \\
AI Literacy--Exposure & --- & --- & .755*** & .308*** & $< .001$ & 4.87 \\
\multicolumn{7}{l}{\emph{Block 4: Expectancy-Value}} \\
Expectancy-Value & --- & --- & --- & .630*** & $< .001$ & 4.87 \\
\multicolumn{7}{l}{\emph{Model Statistics}} \\
$R^2$ & .061 & .330 & .465 & .722 & & \\
Adjusted $R^2$ & .053 & .319 & .455 & .716 & & \\
$\Delta R^2$ & .061 & .264 & .145 & .256 & & \\
F for $\Delta R^2$ & 4.80*** & 143.10*** & 96.76*** & 335.92*** & & \\
\bottomrule
\multicolumn{7}{p{8.2cm}}{\footnotesize\emph{Note.} $\beta$ = standardized regression coefficient. $\Delta R^2$ = increment in $R^2$ from previous block. Reference categories: Gender = non-woman; First-Gen = continuing-generation; STEM = non-STEM; SES = middle/upper. * $p < .05$. ** $p < .01$. *** $p < .001$.} \\
\end{tabular}
\end{table}

The multinomial logistic regression model was statistically significant (LR $\chi^2(21, N = 382) = 164.52$, $p < .001$; McFadden pseudo-$R^2 = .175$; AIC $= 823.44$). AI Literacy was a significant negative predictor across all three contrasts: Instrumental vs.\ Strategic ($OR = 0.37$, $p = .006$), Dependent vs.\ Strategic ($OR = 0.11$, $p < .001$), and Dialogic vs.\ Strategic ($OR = 0.25$, $p < .001$). The Dependent contrast produced the largest effect: each unit increase in AI literacy was associated with approximately a tenfold reduction in the odds of Dependent versus Strategic classification. AI Literacy-Exposure showed the inverse pattern (ORs: 2.89--6.83), with greater prior exposure positively predicting all non-Strategic types. Expectancy-Value beliefs were strong positive predictors of all three non-Strategic types (ORs: 2.07--2.17), with near-identical odds ratios across contrasts. Demographic predictors were non-significant in all contrasts. Table~\ref{tab:multinomial} presents all multinomial logistic regression results.

\begin{table*}
\caption{Multinomial Logistic Regression: Predictors of Dominant Reliance Type (Reference = Strategic, $N = 382$).}
\label{tab:multinomial}
\small
\begin{tabular}{lccc ccc ccc}
\toprule
 & \multicolumn{3}{c}{\textbf{Instrumental vs.\ Strategic}} & \multicolumn{3}{c}{\textbf{Dependent vs.\ Strategic}} & \multicolumn{3}{c}{\textbf{Dialogic vs.\ Strategic}} \\
\cmidrule(lr){2-4}\cmidrule(lr){5-7}\cmidrule(lr){8-10}
\textbf{Predictor} & \textbf{OR} & \textbf{95\% CI} & \textbf{p} & \textbf{OR} & \textbf{95\% CI} & \textbf{p} & \textbf{OR} & \textbf{95\% CI} & \textbf{p} \\
\midrule
Gender (woman) & 0.99 & [0.54, 1.84] & .986 & 0.87 & [0.29, 2.59] & .807 & 0.72 & [0.40, 1.32] & .288 \\
First-generation & 2.16* & [1.04, 4.46] & .038 & 2.32 & [0.64, 8.34] & .198 & 1.61 & [0.77, 3.35] & .205 \\
STEM major & 0.92 & [0.45, 1.86] & .812 & 0.65 & [0.20, 2.13] & .477 & 1.14 & [0.56, 2.32] & .713 \\
Lower SES & 1.72 & [0.85, 3.47] & .129 & 0.51 & [0.13, 2.03] & .337 & 1.47 & [0.73, 2.96] & .276 \\
AI Literacy composite & 0.37** & [0.18, 0.75] & .006 & 0.11*** & [0.03, 0.37] & $< .001$ & 0.25*** & [0.12, 0.50] & $< .001$ \\
AI Literacy--Exposure & 2.89** & [1.45, 5.73] & .002 & 6.83** & [1.96, 23.83] & .003 & 3.29*** & [1.67, 6.49] & $< .001$ \\
Expectancy-Value & 2.17*** & [1.70, 2.77] & $< .001$ & 2.07*** & [1.35, 3.17] & $< .001$ & 2.11*** & [1.66, 2.68] & $< .001$ \\
\bottomrule
\multicolumn{10}{p{16.2cm}}{\footnotesize\emph{Note.} OR = odds ratio. 95\% CIs computed via Wald method. Reference category = Strategic reliance. Dependent group $n = 17$; interpret with caution. * $p < .05$. ** $p < .01$. *** $p < .001$.} \\
\end{tabular}
\end{table*}

The key interpretive insight is that AI literacy and Expectancy-Value beliefs operate at distinct levels: EVT predicts how intensely students rely on LLMs overall, while AI literacy determines which type of reliance they adopt. These are separate predictive systems requiring separate interventions.

\subsection{RQ3: Moderation of Reliance-Outcome Relationships}

\begin{figure}
\centering
\includegraphics[width=\linewidth]{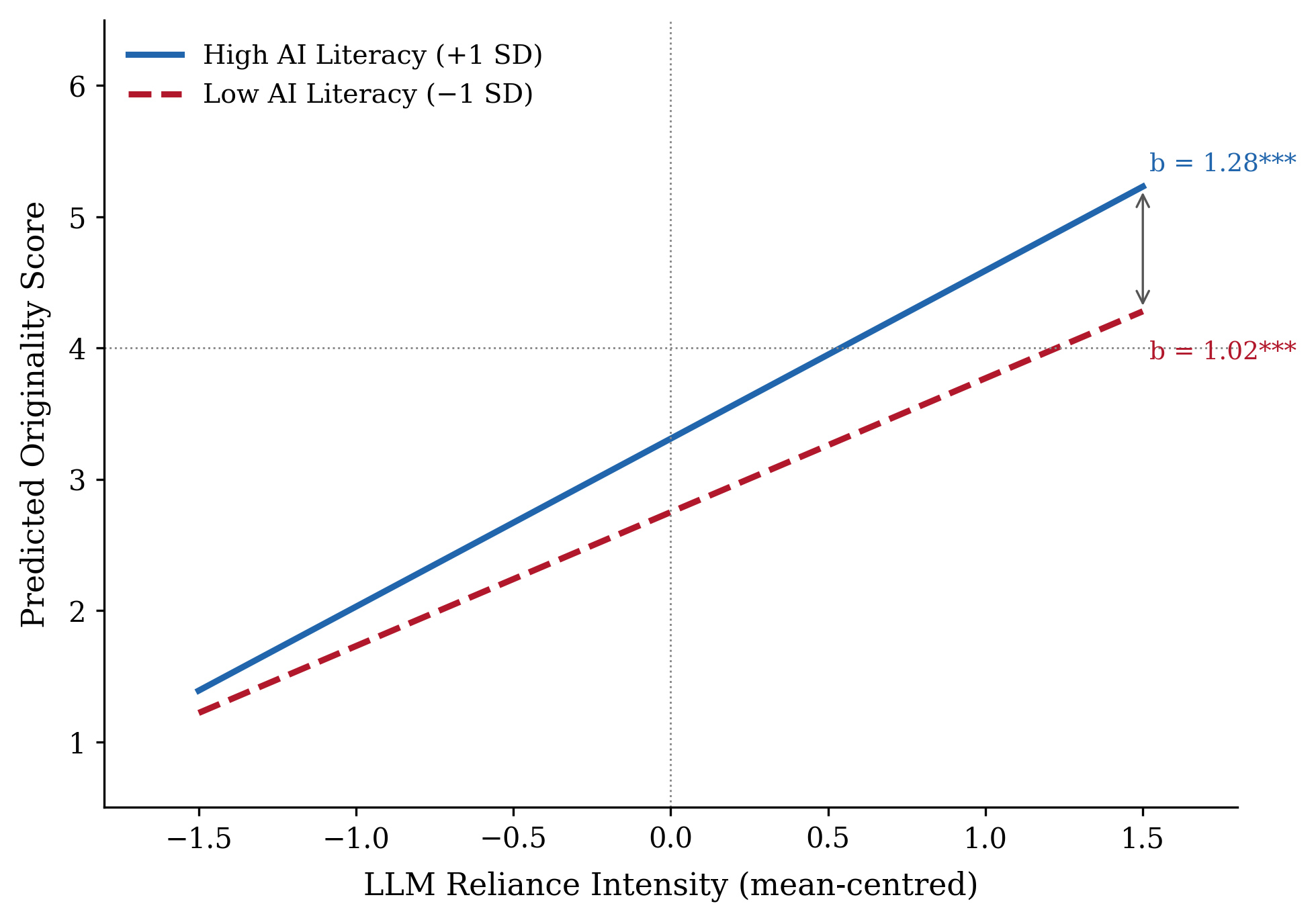}
\caption{H3a: AI Literacy Moderates Reliance Intensity Predicting Originality ($\beta = +.092$, $p = .012$). Simple slopes plotted at $+1$ SD and $-1$ SD of AI Literacy Composite.}
\Description{Simple slopes plot showing predicted originality as a function of LLM reliance intensity at high and low levels of AI literacy. Both slopes are positive, with the high-literacy slope steeper.}
\label{fig:mod-ail}
\end{figure}

\begin{figure}
\centering
\includegraphics[width=\linewidth]{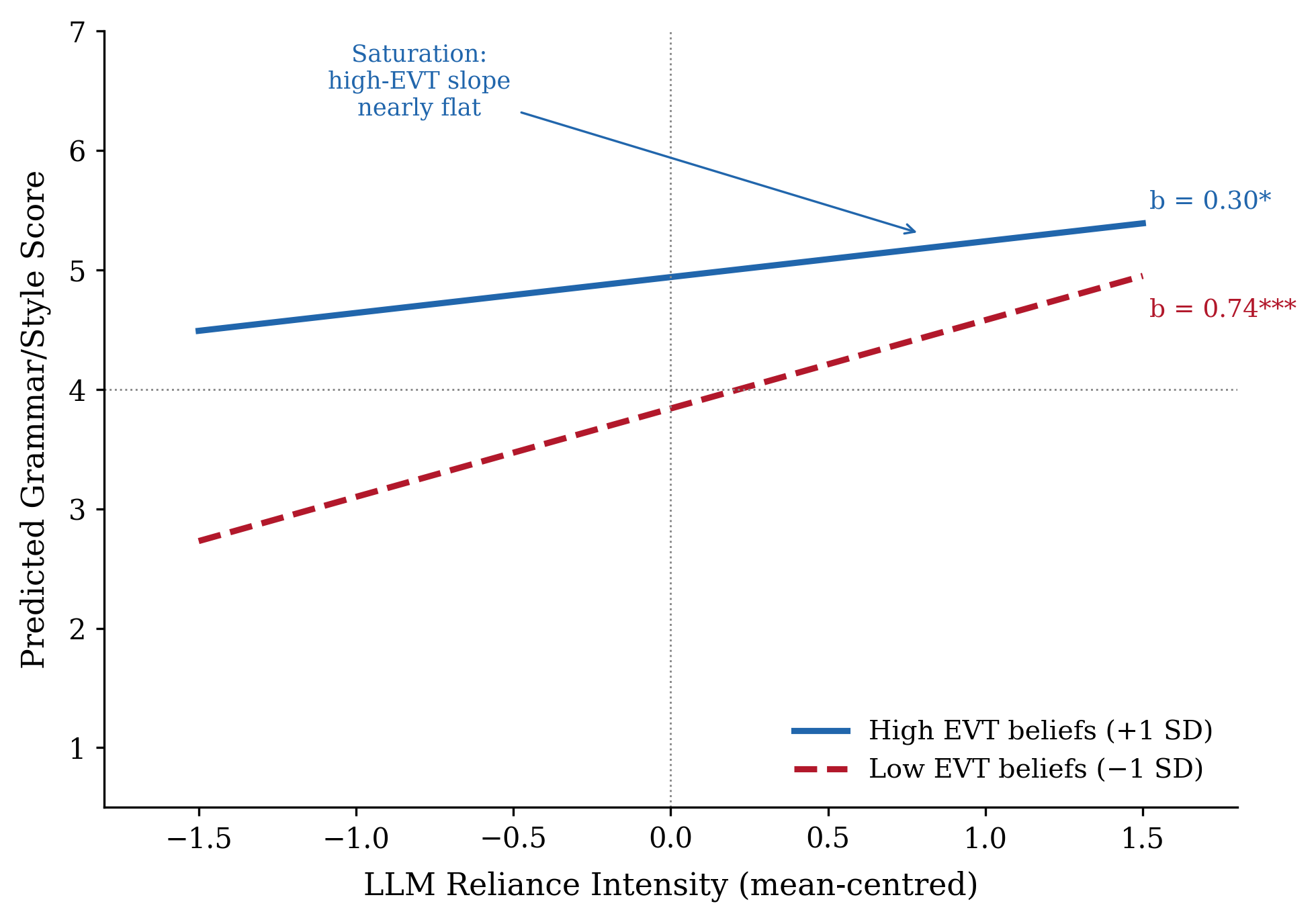}
\caption{H3b: Expectancy-Value Moderates Reliance Intensity Predicting Grammar/Style ($\beta = -.126$, $p = .002$). Simple slopes plotted at $+1$ SD and $-1$ SD of Expectancy-Value.}
\Description{Simple slopes plot showing predicted grammar and style scores as a function of LLM reliance intensity at high and low levels of expectancy-value beliefs. The high-EVT slope is nearly flat, indicating saturation.}
\label{fig:mod-evt}
\end{figure}

Thirteen moderation models were estimated. Six significant interactions emerged. For H3a (AI Literacy as Moderator), two significant interactions were identified: AI Literacy $\times$ Reliance Intensity predicting Originality ($\Delta R^2 = .008$, $F(1, 377) = 6.43$, $p = .012$, $\beta = +.092$), where higher literacy amplified rather than constrained perceived creative benefits, with simple slopes $b = 1.28$ ($+1$ SD) and $b = 1.02$ ($-1$ SD) (both $p < .001$); and Dependent Reliance $\times$ AI Literacy predicting Self-Efficacy ($\Delta R^2 = .007$, $F(1, 378) = 4.94$, $p = .027$, $\beta = -.085$), where higher literacy buffered the self-efficacy inflation associated with dependent use (slopes: $b = 0.71$ at $+1$ SD vs.\ $b = 0.92$ at $-1$ SD). For H3b (EVT as Moderator), EVT significantly moderated Reliance Intensity predicting Grammar/Style ($\Delta R^2 = .014$, $F(1, 376) = 10.16$, $p = .002$, $\beta = -.126$), consistent with a saturation effect (high-EVT slope: $b = 0.30$; low-EVT slope: $b = 0.74$), and Reliance Intensity predicting Originality ($\Delta R^2 = .010$, $F(1, 377) = 7.55$, $p = .006$, $\beta = +.103$). For H3c (Prior Exposure as Moderator), Dependent Reliance $\times$ Exposure predicting Self-Efficacy was significant ($\Delta R^2 = .012$, $F(1, 379) = 8.70$, $p = .003$, $\beta = -.135$), consistent with a developmental recalibration effect. Figures~\ref{fig:mod-ail} and~\ref{fig:mod-evt} plot the simple slopes for the significant AI Literacy and Expectancy-Value interactions, respectively. Table~\ref{tab:moderation} presents the complete moderation summary.

\begin{table*}
\caption{Summary of All Moderation Tests: Reliance Intensity $\times$ Moderator Predicting Writing Outcomes.}
\label{tab:moderation}
\small
\begin{tabular}{llccccl}
\toprule
\textbf{H} & \textbf{Interaction} & \textbf{Outcome} & $\boldsymbol{\Delta R^2}$ & \textbf{F} & \textbf{p} \quad $\boldsymbol{\beta}$ & \textbf{Interpretation} \\
\midrule
H3a & Dep $\times$ AI Literacy & Self-Efficacy & .007 & 4.94 & .027 \quad $-$.085 & Buffering: supported \\
H3a & RI $\times$ AI Literacy & Originality & .008 & 6.43 & .012 \quad $+$.092 & Amplification (unexpected) \\
H3a & RI $\times$ AI Literacy & Quality & .000 & 0.12 & .725 \quad --- & Not significant \\
H3a & RI $\times$ AI Literacy & Self-Efficacy & .001 & 0.61 & .437 \quad --- & Not significant \\
H3a & RI $\times$ AI Literacy & Critical Thinking & .004 & 2.85 & .093 \quad --- & Not significant \\
H3b & RI $\times$ EVT & Grammar/Style & .014 & 10.16 & .002 \quad $-$.126 & Attenuation: supported \\
H3b & RI $\times$ EVT & Originality & .010 & 7.55 & .006 \quad $+$.103 & Amplification (emergent) \\
H3b & Instr $\times$ EVT & Grammar/Style & .009 & 6.54 & .011 \quad $-$.103 & Attenuation replicated \\
H3b & RI $\times$ EVT & Quality & .004 & 3.66 & .057 \quad --- & Not significant \\
H3b & RI $\times$ EVT & Critical Thinking & .001 & 0.64 & .424 \quad --- & Not significant \\
H3c & RI $\times$ Exposure & Self-Efficacy & $< .001$ & 0.07 & .797 \quad --- & Not supported \\
H3c & RI $\times$ Exposure & Originality & .010 & 7.67 & .006 \quad $+$.102 & Amplification (emergent) \\
H3c & Dep $\times$ Exposure & Self-Efficacy & .012 & 8.70 & .003 \quad $-$.135 & Recalibration effect \\
\bottomrule
\multicolumn{7}{p{15.8cm}}{\footnotesize\emph{Note.} RI = overall reliance intensity; Dep = Dependent reliance subscale; Instr = Instrumental reliance subscale; EVT = Expectancy-Value Theory composite; Exp = AI Literacy-Exposure. $\Delta R^2$ represents the incremental variance attributable to the interaction term. $\beta$ = standardized regression coefficient. --- = interaction not significant.} \\
\end{tabular}
\end{table*}

\subsection{Qualitative Findings: Typological Validation and Theoretical Advancement}

A reflexive thematic analysis conducted across three qualitative strands yielded seven distinct themes, comprising 1,435 coded instances. The first theme, Functional Utility (176 instances), elucidates that the use of large language models (LLMs) is both stage-specific and functionally differentiated, with grammar correction and text refinement emerging as the most prevalent applications across diverse demographic groups. The second theme, Reliance Typologies in Practice (120 instances), affirms the validity of the quantitative typology in capturing variation in reliance orientation, while also uncovering a developmental trajectory that is not discernible in cross-sectional analyses. Participants articulated a progression from dependent to strategic reliance over successive semesters, a shift exemplified by Destiny's observation: ``I realized my grades were going up\ldots\ but I wasn't actually learning anything.'' The third theme, Epistemic Risk and Verification (93 instances), corroborates the moderation effect of AI literacy by documenting an almost universal encounter with hallucinated content. All 14 interview participants and 22 of 35 SUR35 respondents reported fabricated citations, with verification practices emerging primarily through self-directed experimentation rather than formalized instruction.

The fourth theme, Motivational Calculus (136 instances), offers qualitative substantiation for the predominance of expectancy-value theory (EVT) ($\Delta R^2 = .256$), with time constraints and task efficiency emerging as the principal motivational determinants of LLM engagement across all three qualitative strands. This pattern was observed among 11 of 14 interview participants, 24 of 35 SUR35 respondents, and 26 of 361 SUR361 respondents. The fifth theme, Ethical Navigation (140 instances), reveals that instructor policy functions as an externalized moral reference for the majority of students. Notably, 11 of 14 interview participants and 18 of 35 SUR35 respondents identified the AI policy articulated by their instructors as the primary factor shaping their decisions regarding LLM use in each course.

The seventh theme, Critical AI Ethics, Environmental Concerns, and Principled Non-Use (188 instances across strands), constitutes the study's most significant theoretical extension. Within the full SUR361 strand, 45 respondents (12.5\%) spontaneously referenced environmental concerns, 47 (13.0\%) articulated principled non-use rooted in personal ethical commitments, and 12 (3.3\%) expressed apprehensions regarding cognitive offloading. These findings necessitate the articulation of a Two-Model Architecture: the four-type reliance taxonomy and its EVT-based predictor model adequately characterize negotiation-mode students, who weigh expected value against cognitive cost in formulating context-sensitive reliance decisions. In contrast, abstention-mode students, principled non-users whose prior ethical commitments serve as categorical constraints, are not encompassed by extant reliance frameworks. Their consistently low scores on reliance measures are attributable not to limited AI literacy or attenuated EVT beliefs, but rather to a values-based categorical refusal, a motivational architecture that lies entirely beyond the explanatory scope of the EVT framework.

\section{Discussion}

The findings of this study advance a central argument that both extends and critically revises the prevailing literature on large language model (LLM) reliance: namely, that the field has been engaged in measuring constructs that do not adequately capture the intended phenomena. By operationalizing LLM outcomes through perceived AI-mediated attainment items, prior scholarship, including the present study prior to qualitative reinterpretation, has systematically mischaracterized the most sophisticated students as the least effective. Those students who exhibit advanced AI literacy, adhere to principled writing practices, and exercise genuine metacognitive oversight in their engagement with AI are consistently evaluated as less effective by outcome measures that privilege the extent of AI contribution to writing. For these individuals, minimal AI involvement constitutes the most principled approach. This observation is of considerable significance. The current body of evidence concerning `outcomes' in LLM reliance research thus primarily reflects the degree of AI throughput and, at its most problematic, risks incentivizing cognitive dependency at the expense of intellectual authorship.

\subsection{The Prediction Architecture: Two Levers, Two Interventions}

Hierarchical regression and multinomial logistic regression analyses identified two empirically distinct predictor systems. Expectancy--Value Theory (EVT) beliefs emerged as the primary predictor of reliance intensity ($\beta = .630$, $\Delta R^2 = .256$, total $R^2 = .722$), explaining more variance than demographics, prior exposure, and AI literacy combined. In contrast, AI literacy predicted reliance \emph{type}, reflecting the qualitative nature of engagement rather than its magnitude ($OR = 0.11$, Dependent vs.\ Strategic, $p < .001$). These systems are not redundant; for example, students with high EVT beliefs may exhibit any of the four reliance types, while those with high AI literacy may still engage.

The practical implication is clear: AI literacy interventions and motivational interventions function as distinct levers. Programs that enhance students' understanding of large language model (LLM) capabilities and limitations, their ability to evaluate AI-generated outputs, and their ethical awareness of AI use can promote a shift from uncritical to critical engagement, but may not reduce the overall volume of engagement. Interventions focused solely on knowledge and skills, without addressing motivational beliefs, perceived utility, efficiency value, and expectancy for success, will influence only one dimension of reliance while leaving the other unaffected.

This two-lever framework aligns with recent empirical findings. \citet{chan2023} demonstrated that perceived value, rather than perceived cost, was the primary determinant of students' intentions to use generative AI in an EVT-based assessment, indicating that motivational architecture shapes the breadth of engagement. \citet{tulis2025} reported that motivational appraisals, particularly challenge framing, accounted for significant variance in generative AI engagement beyond the cognitive predictors identified by the Unified Theory of Acceptance and Use of Technology (UTAUT). Collectively, these findings suggest that AI literacy curriculum design should position motivational beliefs as an instructional target equal in importance to knowledge and skills, rather than treating them as a secondary consideration.

\subsection{The Measurement Artifact: Field-Wide Implications}

The most theoretically significant quantitative finding reveals that Strategic users, defined by their deliberate, metacognitively regulated, and academically principled approach, scored lowest across all ten writing-process engagement and outcome variables (all ANOVA $p$s $< .001$; $\eta^2$ range $= .166$--$.329$). Notably, the contrast between Strategic and Dependent users produced the largest effect sizes observed in the study (Planning: $d = 1.61$; Quality: $d = 1.73$), demonstrating that the most autonomous and the most dependent writers differed by nearly two standard deviations on outcomes that, in principle, would be expected to favor the autonomous group.

Qualitative evidence indicates that this pattern is best understood as a measurement artifact rather than a reflection of genuine performance differences. All outcome items in this study, as well as those in the reviewed LLM reliance instruments, are structured as statements such as ``I use generative AI tools to achieve X.'' Such items primarily assess the degree to which students attribute their writing outcomes to AI engagement, rather than directly measuring writing quality, cognitive development, or academic achievement. Consequently, students who deliberately limit AI involvement in their work inevitably score low on these items, not because of deficiencies in their writing, but as a result of their principled restraint, which leads to low AI-attributed attainment. The case of Tyler, the Strategic user whose profile most clearly exemplified this phenomenon, is illustrative: his approach, described as ``trying to disprove it almost, so I'll go back to the textbook,'' resulted in a near-zero Originality score, reflecting minimal AI contribution rather than a lack of originality in his work.

This finding carries significant implications for the broader field, extending well beyond the scope of the present study. As \citeauthor{lintner2024}'s \citeyearpar{lintner2024} systematic review demonstrates, only a minority of the sixteen validated AI literacy scales have been evaluated for responsiveness, understood as the capacity to detect genuine change in the construct over time. Moreover, none have been assessed for cross-cultural validity or measurement error, which raises a critical validity concern: the field currently lacks outcome instruments capable of distinguishing between AI-assisted and AI-independent performance, as well as predictor instruments validated against observable behavioral change. Until these methodological gaps are addressed, the evidence base concerning effective outcomes among LLM users will remain systematically confounded by the limitations inherent in existing measurement frameworks.

\subsection{The Two-Model Architecture: A Necessary Structural Extension}

A reflexive thematic analysis of the SUR361 strand identifies a population not accounted for by the prevailing four-type quantitative typology. Specifically, 13.0\% of respondents to the open-ended survey questions expressed a principled non-use of AI, grounded in ethical, environmental, or epistemic values. These individuals used language indicating a categorical refusal to engage with AI, rather than exhibiting selective restraint. Quantitatively, this group records near-zero scores across all four reliance subscales, a profile that superficially resembles low-intensity Strategic users. However, qualitative analysis indicates that these two groups are characterized by fundamentally distinct motivational structures.

Three diagnostic dimensions distinguish principled non-users from highly Strategic users. First, \emph{directionality of reasoning}: Strategic users engage in ongoing negotiation, determining when and how to use AI based on task demands and authorial goals. Their low scores reflect selective, bounded engagement. In contrast, principled non-users have decided not to use AI, and their near-zero scores reflect categorical exclusion from a decision they consider already settled. Second, \emph{context-sensitivity}: Strategic users' AI engagement varies across task types, assignment demands, and instructor policies, whereas principled non-users' abstention is context-independent, applying the same categorical refusal regardless of task, policy, or time pressure. Third, \emph{motivational architecture}: Strategic users operate within the expectancy-value theory (EVT) negotiation framework, weighing expected value against perceived cost before each engagement decision. Principled non-users, however, resolve the engagement question before any cost--benefit calculation, guided by prior ethical commitments that function as categorical constraints. In summary, Strategic users choose not to rely heavily on AI in specific instances, while principled non-users have chosen not to use AI at all.

Combining these two populations in quantitative instruments introduces heterogeneity into the predictor structure that regression models cannot resolve. When abstention-mode students, whose near-zero scores result from ethical commitments, are grouped with Strategic users, whose restrained scores reflect metacognitive governance, the predictor structure becomes internally inconsistent. The expectancy-value theory (EVT) negotiation model predicts engagement under varying motivational conditions with substantial explanatory power (total $R^2 = .722$), but it lacks a mechanism to account for outcomes determined by categorical prior commitments rather than contextual value calculations. This structural gap necessitates a Two-Model Architecture: a negotiation model for the four-type taxonomy applicable to students for whom engagement remains a behavioral option, and a separate abstention model for students whose prior ethical commitments have already resolved the engagement question categorically. Future scale development should distinguish these populations before incorporating near-zero reliance scores as outcome data in regression models.

\subsection{The MSI Context: Equity as a Foreground Concern}

The institutional context of this study serves as a central analytic lens for interpreting the findings. First-generation status was the only individually significant demographic predictor of reliance intensity in Block 1 ($\beta = .153$, $p = .008$), and its interpretive significance is amplified by the institutional environment. At this institution, approximately one-third of undergraduates are first-generation college students, over a quarter identify as Black or African American, and more than half receive Pell Grant support \citep{oir2025}. These characteristics are not peripheral to the sample; instead, they constitute the structural conditions that define the Minority-Serving Institution (MSI) context in which patterns of AI reliance must be understood.

This interpretation is more consistent with a compensation mechanism than with a deficit in AI literacy. Approximately one-third of students at MSIs are first-generation, compared to about one in five at non-MSIs \citep{pnpi2025}. Therefore, the observed pattern, first-generation students demonstrating higher reliance intensity as a substitute for otherwise inaccessible academic support, should be viewed as a structural feature of the MSI context rather than an individual anomaly. The increased reliance intensity among first-generation students does not necessarily indicate weaker AI literacy; instead, it may suggest that AI functions as a cost-effective alternative to academic support forms, such as family cultural capital, familiarity with writing-center resources, and prior academic preparation, which are more accessible to continuing-generation peers.

This finding builds upon \citeauthor{cotton2024}'s \citeyearpar{cotton2024} research on technology adoption at under-resourced institutions by explicitly situating the compensation mechanism within an Expectancy-Value Theory (EVT) framework. First-generation students' higher reliance intensity may reflect not weaker AI literacy, but a stronger perceived utility value in contexts where alternative academic support is structurally limited. The equity implications are significant. If AI literacy programs redirect first-generation students from uncritical to strategic reliance without addressing the underlying resource gap that drives higher intensity, these programs alter student engagement without changing the structural conditions that shape such engagement.

At Minority-Serving Institutions, three intersecting forces produce a distinct and addressable form of educational harm: disparities in AI literacy amplify existing differences in academic preparation; institutional policy frameworks emphasize detection rather than developmental support; and prevailing outcome measures incorrectly classify the most sophisticated students as the weakest performers. The Reliance Negotiation Framework, the Two-Model Architecture, and the measurement-artifact finding presented in this study collectively provide the theoretical tools necessary to address this harm. Whether institutions will implement these concepts remains uncertain, but the urgency of this question is underscored by the present study.

\subsection{Limitations}

The interpretive scope of these findings is circumscribed by several notable limitations. Foremost among these is the cross-sectional design, which necessarily precludes any claims of causal inference. As such, the directionality of the observed relationship between EVT-reliance and, more specifically, the association between AI literacy-type membership, remains indeterminate in the absence of longitudinal data. While the qualitative accounts provided by students, which describe a progression from dependent to more strategic forms of reliance across academic semesters, offer a theoretically persuasive narrative, these developmental trajectories remain empirically unsubstantiated within the methodological constraints of the present study.

The methodological reliance on self-report data introduces a set of social desirability concerns that are especially pronounced in this domain. Notably, the tendency among participants to attribute dependent use to their past selves or to other students during interviews suggests a systematic underreporting of habituated reliance, a phenomenon that has been documented in prior research on self-disclosure in academically sensitive contexts. Furthermore, the present study did not compute formal item-level content validity indices. It is therefore incumbent upon future research to employ established procedures, such as the \citet{lynn1986} method, utilizing a minimum of five expert raters and adhering to an I-CVI threshold of .78, to ensure the robustness of instrument validation.

The generalizability of these findings is further constrained by the single-institution design. The specific demographic composition, R1 classification, and institutional AI policies of the participating university may not be representative of other Minority Serving Institution (MSI) contexts. Consequently, these results must be interpreted with careful attention to the distinctive structural conditions of a public R1 AANAPISI situated in an urban mid-Atlantic environment. Additionally, the moderation power analysis indicated a required sample size of at least 395 to detect small interaction effects ($f^2 = .02$), whereas the actual sample size of 382 fell marginally short of this threshold. As such, non-significant moderation results should be regarded as inconclusive, rather than as definitive evidence of null effects.

A further limitation concerns the measurement-construct gap identified in this study, namely, the inability of extant outcome instruments to adequately distinguish between AI-assisted attainment and independent writing quality. This issue remains unresolved within the present research and constitutes a pressing instrumentation challenge for future investigations into LLM reliance. Addressing this gap will necessitate not only the development of novel item structures but also a more precise theoretical articulation of the intended constructs that AI outcome measures are designed to capture.

\subsection{Implications and Future Directions}

The findings yield four principal implications. First, the design of AI literacy curricula ought to be disentangled from the objective of reducing reliance intensity as a pedagogical end. The evidence indicates that AI literacy is predictive of the mode of reliance rather than its magnitude; thus, curricular interventions that cultivate students' evaluative and ethical capacities are likely to foster a transition from uncritical to critical engagement, even as the overall volume of engagement remains unchanged. This constitutes the appropriate pedagogical aim, yet it necessitates a clear recognition that the development of literacy is not equivalent to a diminution of use. Indeed, interventions oriented toward reducing the frequency of AI use may inadvertently curtail the very forms of engagement that, when properly regulated, serve to advance rather than impede intellectual growth.

Second, expectancy-value theory (EVT) beliefs should be foregrounded as a central target within AI literacy curricula, rather than relegated to the status of a background variable. The data demonstrate that the principal determinant of reliance intensity is motivational rather than competency-based. Consequently, curricular interventions that focus exclusively on the transmission of knowledge and skills, without directly addressing students' beliefs regarding the utility and cost of AI, or the alignment of AI use with their scholarly identities and aspirations, are unlikely to influence the most salient behavioral determinant identified in this research.

Third, there is a pressing need for the comprehensive redesign of outcome measurement instruments in this domain. The field requires tools capable of independently assessing AI-free writing competence subsequent to AI-assisted practice, as well as portfolios of unaided performance, pre- and post-transfer assessments, and longitudinal measures of writing development. Reliance on instruments that merely capture perceived AI-mediated attainment at a single temporal juncture is insufficient. The systematic review substantiates that rigorously validated instruments measuring students' beliefs, perceptions, and practices in AI-assisted academic writing remain notably underdeveloped, with the majority of extant studies employing adapted tools not originally conceived for the generative AI context.

Fourth, future research should prioritize the development and validation of abstention measures that can differentiate among the various rationales underlying near-zero reliance scores, including strategic metacognitive restraint, categorical ethical refusal, low-exposure non-use, and accommodations related to disability. Such differentiation is essential prior to the inclusion of these populations in regression predictor models. The indiscriminate aggregation of abstention-mode students with those exercising strategic restraint conflates motivational structures that are theoretically distinct and that, in practice, necessitate divergent pedagogical interventions.

\section{Conclusion}

This study demonstrates that reliance on large language models (LLMs) in undergraduate academic writing is not a singular phenomenon but is structured into four qualitatively distinct types: Strategic, Instrumental, Dialogic, and Dependent. Each type is characterized by a specific relationship to the affordances of artificial intelligence, produces a unique profile of writing process engagement, and requires a distinct pedagogical response. The predictive architecture underlying these types is two-tiered and empirically separable: expectancy-value theory (EVT) beliefs determine the intensity of LLM reliance, while AI literacy determines the direction, or type, of reliance adopted by students. These predictors are neither redundant nor interchangeable; instead, they represent separate systems that govern distinct dimensions of LLM engagement. Therefore, interventions that address only one dimension while neglecting the other are likely to yield partial and predictably insufficient outcomes.

A key theoretical finding is that Strategic users consistently score lowest on all outcome measures. This result does not provide empirical insight into writing quality; instead, it highlights a measurement issue that challenges the construct validity of the existing literature on LLM outcomes. The outcome instruments currently used in the field, which are structured around perceived AI-mediated attainment, do not measure what they claim to assess. Rather, they primarily capture AI throughput. Without clearer instrumentation for the constructs that AI outcome measures actually assess, the evidence base concerning the consequences of LLM engagement for student learning remains weighted toward rewarding dependence and penalizing the principled restraint essential for genuine intellectual development.

The Minority-Serving Institution (MSI) context does not merely provide demographic background for these findings; it clarifies their implications for educational equity. At institutions where a significant proportion of students are first-generation and the majority receive federal financial aid, the structural conditions that foster compensatory AI reliance are central to the research problem. Three forces converge at MSIs to produce a specific and remediable form of educational harm: disparities in AI literacy exacerbate existing differences in academic preparation, institutional policy frameworks prioritize detection over development, and outcome measures inaccurately portray the most sophisticated students as the weakest performers. The Reliance Negotiation Framework, the Two-Model Architecture, and the Three-Tier Ethical Reasoning Model introduced in this study provide the theoretical tools necessary to address these challenges. Whether institutions will act upon this theoretical foundation remains uncertain, but the urgency of such action is underscored by the present findings.

\bibliographystyle{ACM-Reference-Format}
\bibliography{refs}

\end{document}